\RequirePackage{rotating}
\documentclass[10pt,a4paper,reqno]{amsproc}
\usepackage[utf8]{inputenc}
\usepackage[T1]{fontenc}
\usepackage{enumitem}
\usepackage[foot]{amsaddr}
\usepackage{appendix}
\usepackage[hmargin=3.5cm,vmargin=3cm]{geometry}

\usepackage[mathscr]{eucal}

\usepackage{fancyhdr}
\usepackage[authoryear]{natbib}

\makeatletter
\newtheoremstyle{mytheorem}{1pc}{1pc}{\itshape}{}{\bfseries}{.}{0.3em}{{\thmname{#1}\thmnumber{ #2}\thmnote{ (\mdseries #3)}}}
\makeatother
\makeatletter
\newtheoremstyle{mydefinition}{1pc}{1pc}{}{}{\bfseries}{.}{3pt}{{\thmname{#1}\thmnumber{ #2}\thmnote{ (\mdseries #3)}}}
\makeatother



\renewcommand{\leq}{\leqslant}
\renewcommand{\geq}{\geqslant}

\numberwithin{equation}{section}

\theoremstyle{mytheorem}

\theoremstyle{mydefinition}

\usepackage{amsmath}
\usepackage{amssymb}
\usepackage{wasysym}
\usepackage{tikz}

\usepackage{bm}
\newcommand{\R}{\mathbb{R}}
\newcommand{\N}{\mathbb{N}}
\newcommand{\x}{\mathbf{x}}
\newcommand{\XX}{\overrightarrow{\x}}
\newcommand{\pp}{\bm{\pi}}
\newcommand{\Sr}{\mathbf{\mathcal{S}}_r}
\newcommand{\1}{\mathbf{1}}
\newcommand{\LL}{\mathrm{L}}
\newcommand{\dd}{\mathrm{d}}
\newcommand{\K}{\mathrm{K}}

\renewcommand{\t}{\mathbf{t}}
\renewcommand{\tt}{\overrightarrow{\t}}

\usepackage[modulo]{lineno}
\begin{document}

%%%%%%%%%%%%%%%%%%%%%%%%%%%%%%%%%%%%%%%%%%%%%%%%%%%%%%%%%%%%%%%%%%%%%%%%%%%%%%%
%%% Spaces around equations and between equation array lines
%%%%%%%%%%%%%%%%%%%%%%%%%%%%%%%%%%%%%%%%%%%%%%%%%%%%%%%%%%%%%%%%%%%%%%%%%%%%%%%

%\setlength{\abovedisplayskip}{1ex plus .2ex minus .2ex}
%\setlength{\belowdisplayskip}{1ex plus .2ex minus .2ex}
%\setlength{\jot}{0.5ex}

%%%%%%%%%%%%%%%%%%%%%%%%%%%%%%%%%%%%%%%%%%%%%%%%%%%%%%%%%%%%%%%%%%%%%%%%%%%%%%%
%%% Title stuff
%%%%%%%%%%%%%%%%%%%%%%%%%%%%%%%%%%%%%%%%%%%%%%%%%%%%%%%%%%%%%%%%%%%%%%%%%%%%%%%
 
\title[SSPP Models for Forest Tree Data]{Sequential Spatial Point Process Models for Spatio-Temporal Point Processes: A Self-Interactive Model with Application to Forest Tree Data }

\author[Yazigi, A.]{\bf Adil Yazigi}
\address{Adil Yazigi\\
School of Computing, University of Eastern Finland, Joensuu, Finland}
\email{adil.yazigi@uef.fi}

\author[Penttinen, A]{\bf Antti Penttinen}
\address{Antti Penttinen\\
	Department of Mathematics and Statistics, University of Jyv\"{a}skyl\"{a}, Finland}
\email{antti.k.penttinen@jyu.fi}

\author[Ylitalo, A.-K.]{\bf Anna-Kaisa Ylitalo}
\address{Anna-Kaisa Ylitalo\\
	Natural Resources Institute Finland (Luke), Finland}
\email{anna-kaisa.ylitalo@luke.fi}

\author[Maltamo, M]{\bf Matti Maltamo}
\address{Matti Maltamo\\
	School of Forest Sciences, University of Eastern Finland, Joensuu, Finland}
\email{matti.maltamo@uef.fi}

\author[Packalen, P.]{\bf Petteri Packalen}
\address{Petteri Packalen\\
School of Forest Sciences, University of Eastern Finland, Joensuu, Finland}
\email{petteri.packalen@uef.fi}

\author[Meht\"{a}talo, L.]{\bf Lauri Meht\"{a}talo}
\address{Lauri Meht\"{a}talo\\
	School of Computing, University of Eastern Finland, Joensuu, Finland}
\email{lauri.mehtatalo@uef.fi}

%\thanks{A. Yazigi was funded by the Finnish Doctoral Programme
%in Stochastics and Statistics.}

%\subjclass[]{}

%\keywords{Marginal distribution; Maximum likelihood;  Self-interaction; Sequential spatial point processes;  Spatio-temporal point processes.}

\vspace{18mm} \setcounter{page}{1} \thispagestyle{empty}

\begin{abstract}
We  model the spatial dynamics of a forest stand by using a special class of spatio-temporal point processes, the sequential spatial point process,   where  the spatial dimension is parameterized and the time component is atomic.  The sequential spatial point  processes differ from spatial point processes in the sense that the realizations are ordered sequences of spatial locations and the order of points  allows us to approximate the spatial evolutionary dynamics of the process. This feature shall be useful to interpret the long-term dependence and the  memory formed by the spatial history of the process.  As an illustration, the sequence can represent the tree locations ordered  with respect to time, or to some given quantitative marks  such as tree diameters. We derive a parametric sequential spatial point process model that is expressed in terms of self-interaction of the  spatial points, and then the  maximum-likelihood-based inference is tractable. As an application, we apply the model obtained to forest dataset collected from the Kiihtelysvaara site in Eastern Finland. Potential applications   in remote sensing of forests are discussed.
\end{abstract}

%\today

\maketitle

%{\small
%\noindent\textbf{Mathematics Subject Classification (2010):} 
%60G15, 60G22, 91G80.

\noindent\textsc{\small Keywords}: \small  Marginal distribution; Maximum likelihood;  Self-interaction; Sequential spatial point processes;  Spatio-temporal point processes.

%%%%%%%%%%%%%%%%%%%%%%%%%%%%%%%%%%%%%%%%%%%%%%%%%%%%%%%
%%% THE BEEF
%%%%%%%%%%%%%%%%%%%%%%%%%%%%%%%%%%%%%%%%%%%%%%%%%%%%%%%

%%%%%%%%%%%%%%%%%%%%%%%%%%%%%%%%%%%%%%%%%%%%%%%%%%%%%%%
\section{Introduction}
\label{s:intro}
%%%%%%%%%%%%%%%%%%%%%%%%%%%%%%%%%%%%%%%%%%%%%%%%%%%%%%%

%Here is the introduction. The file \texttt{biomsample.tex} gives an
%example of using the \texttt{natbib} package to cross-reference
%citations from the bibliography.  Here, we just do this manually to
%demonstrate that the old-fashioned way also is acceptable.  See the
%comment right before the bibliography section below for information on
%using BiBTeX.
%
%Please note that, although this document class produces a final
%product that is very close to the format and appearance of a typeset
%\textit{Biometrics} article, there may be some idiosyncracies that
%cause the format to deviate slightly from that in the journal.  These
%will be corrected at the typesetting phase should your paper be accepted
%and forwarded for publication.  So do not worry if such things occur!

Spatial point patterns are often seen as designs unveiling the spatial structures observed in many fields such as epidemiology, seismology, image analysis and forestry. The common characteristic in these types of patterns  is that the most interesting variable to be analyzed is the location of an event.  Sometimes, the locations are endowed with quantitative marks or attributes carrying additional information  such as  the sizes of  trees in  forest stand dataset, or the magnitude and the depth for earthquakes.
%, so we may extend the class of SPP's to be of marked point processes (MPP's).  

However, analyses carried out by spatial point process models
%, as well as MPP's,
have mainly focused on addressing  point patterns within a purely spatial framework, where we thoroughly ignore the underlying evolutionary dynamics in which the occurring instances of events
%, or marks,
are fundamentally time dependent. Here  one should consider the approach of spatio-temporal point processes (STPP's)
%, or spatio-temporal marked point processes(STMPP),
instead, once the temporal information is taking place. Roughly speaking,  a STPP is a collection of instantaneous events, each occurring at
a given spatial location $x_i$, with a given associated time  $t_i$,
% random sequence of time instances associated with spatial locations,
namely, $N_{st}= \{(x_i, t_i), i= 1, \ldots, n\}$ for an integer $n \in \N_0$. For  more details on STPP models we refer to  \citet{cressie1993,daleyverejones2008,diggle2013},  and  to a  review by \citet{gonzalez2016}. 

The temporal aspect within the structure of a STPP offers a natural ordering of the points, which does not exist in spatial dimension. In fact, one may often think  of STPP as a purely temporal point process where each point is associated  with a spatial mark. On the other hand, the temporal dimension under this construction reveals the evolutionary character established by the accumulated information and data generating mechanism built upon the ordered sequence of times $(t_1, \ldots, t_n)$, which is a point process itself. Also, the likelihood can be computed sequentially. 
%In this case, all the information is collected into the internal history to which the process is adapted, and  the density function at a given $t_i$ can be evaluated a as function of the past history formed by  $t_j, j<i.$ 

Nevertheless,  questions arises on the importance of the spatial dimension in this case, and on whether the  dynamics of the observed phenomena could be described by the spatial marginal distribution without any loss of information. Addressing these issues leads to  a parameterization of  events  $\{x_i\}$, which can be examined as a realization of \textit{sequential spatial point process} (SSPP), see \citet{lieshout2006a,lieshout2006b}, and \citet{lieshout2009}. Under this sequential construction, a realization of the SSPP is an ordered sequence  $\XX_n=(x_1, \ldots, x_n)$ of spatial locations where the time component is disclosed as an auxiliary information. 

Following the general framework in \citet{lieshout2006a}, a  SSPP can be identified with a  vector of  ordered  points of a STPP. This also holds for the case where each point is attached to a mark.  Conversely,  the joint distribution of a STPP can be decomposed into  conditional and marginal distributions, which we describe heuristically as
\begin{linenomath} 
	\begin{equation*}
	[\mathrm{S},\mathrm{T}] = [\mathrm{S}]_{\text{\scriptsize  ordered}}\,  \cdot [\mathrm{T}|\mathrm{S}],
	\end{equation*}
\end{linenomath} 
where $\mathrm{T}$ indicates the time and $\mathrm{S}$ the spatial component of the process. The spatial marginal distribution inherits the time-order and defines a SSPP  which we denote by $N_{ss}= \{\XX_n\}$, with $\XX_0= \emptyset.$ Moreover, the distribution at  a given location $x_i \in \XX_n$ is a function of the past history, therefore, this enables the SSPP model to capture  the  built-up information through the sequence,  reflecting the long-term spatial dependence between ordered locations and the spatial memory formed during the evolution of the process.

% Conversely, following the general framework in van Lieshout (2006a), a (marked) SSPP can be identified with a  vector of an ordered (marked) points of a (marked) STPP. 

Other constructions of SSPP can be found in the works of \citet{evans1993} and \citet{tablot},  known as random sequential absorption models, or similarly as simple sequential inhibition by \citet{diggle1976} and \citet{lieshout2006b,lieshout2006c}. Recently, significant generalizations and extensions of the SSPP models have been proposed for ordered spatial point patterns by \citet{penttinen2016} in studying  eye movement, and by \citet{moller2016} in modeling forest stand data by transforming a marked point process into a STPP.

In this paper,  we are mainly concerned with forest characteristics. Naturally, forest stands can be thought as realizations of  a spatio-temporal point process. In ordered point patterns, tree size is a natural surrogate for time as large trees are usually older than the smaller trees \citep{moller2016}. Therefore, the locations of large trees have an effect on the locations of small trees, but not necessary the opposite. The motivation for our model development is in airborne laser scanning based forest inventory \citep[see e.g.][]{maltamo2014}. In addition,  our suggested SSPP model is also  useful in modeling inter-tree competition and revealing forest stand dynamics that is not present in static point patterns. 

In airborne laser scanning (ALS), a lidar device carried by aircraft takes repeated height measurements of the area below. ALS produces a set of echoes (points) which is not regular but dense, e.g. many echoes per meter square. Individual trees can be detected from the ALS data with several methods. However, a common problem is that small trees growing below the canopies of bigger trees are  often hidden and mean tree size is overestimated, and  therefore,  the intensity of the point process (stand density) is  underestimated. The hidden trees can be taken into account in estimation of stand density by modeling the detectability of trees as a function of tree size to adjust the observed trees by the estimated detectability  using a Horvitz-Thompson-like estimator \citep{kasper2016,lauri2006}. For this adjustment, we need an evolutionary model for the forest to estimate the relative intensity of the small trees within the influence zone of larger trees compared to non-influenced areas. \citet{kasper2016} assumed that relative intensity is similar in influenced and non-influenced areas.

We shall derive a parametric SSPP model that is  \textit{self-interactive}, and where computing the likelihood is straightforward. Following the terminology in \citet{penttinen2016}, the model enjoys the self-interaction feature when given  past observations $x_1, \ldots, x_k$,  the information carried by a new point $x_{k+1}$ 
%the effected move from a point $x_k$ to the next point $x_{k+1}$
is accommodated by the model, which alters the probability law.
%whenever the move $x_k \to x_{k+1}$ is made.
In fact, this feature eventually expresses the long-term spatial dependence between the trees and the effects of the  memory in the SSPP model. Our motivation  here is  to see whether employing the mechanism of the SSPP model with the self-interaction property would indicate any presence of location-dependence  among ordered sequence of trees, assessed by the parameters of the self-interaction function. The model is fitted to real forest data of 79 plots, but for illustration, we investigate two plots with different spatial structures in order to demonstrate the potential of the suggested model. Plots include tree locations attached with quantitative marks representing the diameter of the trees at breast height (DBH).  In particular, the arrangement of the sequence will be taken in  a descending order  expressed in terms of DBH, i.e. taking the largest tree  as the first event and the smallest tree as the last event in the sequence. Self-interaction  parameters for the model will be estimated by using  the maximum likelihood method once the likelihood is obtained. To evaluate the model, several summary  statistics assisted by Monte Carlo simulation are applied in model evaluation. Here we work with summary statistics that measure different features of the data, and that  take the temporal order into account.
%         of a given point with the past, changes in the inter-distances between consecutive points relative to the closest
%           recurrence, interaction-area, inter-distance and ball union coverage as  functions of time (or order).

The outline of the paper is as follows: In Section 2 we discuss the sequential approach for the spatial point processes and introduce the proposed SSPP model, while the statistical inference for the  obtained model  is   treated in Section 3. Section 4 presents  simulation experiments demonstrating  the model and explores the proposed summary statistics.  In Section 5 we fit the model to the tree  dataset, and  Section 5  presents final remarks and discussion for future work. 

\section{The Finite Sequential Spatial Point Process Models}
\label{s:model}

\subsection{Background}\label{prelim}

%The Cox model (Cox, 1972) is one of the most widely used statistical
%models.  Hastie, Tibshirani, and Friedman (2001) is an example of a
%citation to a work with three authors.  The first time you reference
%one of these in the text, use all the authors names. However, in all
%subsequent references, just use Hastie et al. (2001).  Works with four
%or more authors are always referenced in the text using ``et al.''
%All authors names should appear in the bibliography for all entries.
%
%Please use a recent issue of \textit{Biometrics} as a guide to the style
%for citations and bibliography entries, and follow that style exactly!!

%On a bounded region in the plane, let the observed data $(x_i, m_i),  i=1, \ldots, n$ with finite $ n \in \N_0 $, be a set of points $x_i$ indicating the locations of trees and a set of corresponding marks $m_i \in \R$ representing the sizes of the trees which are the DBH's, and such that the following order  $0< m_1 < m_2< \ldots <m_n < \infty $ holds. 
%%We refer to this  ordered sequence of marks by $\mmm=(m_1, \ldots, m_2) $.
%% As one could arranges the marks 
%The data can be viewed as a realization of a STPP where the occurrence times are $t_i=m_i$. In fact, one could consider a marked point process instead
%
% The data can be viewed as a realization of a marked point process (MPP), or  of a spatio-temporal point process by taking the  times $t_i = m_i$ for all $i$. However, there is a one-to-one correspondence between the class of MPP's and the class of STPP's, see  M\o{}ller et al. (2016). In this paper we 

We consider a finite STPP whose realizations are consisting of distinct points $(t_i,x_i), \text{ for } i=1, \ldots, n, n \in N_0,$ in a compact spatial domain $D \subset \R^2$ and time interval $T \subset R$.

%  We note here that in the case of a  spatial point process with a random quantitative mark, the process can be transformed into a STPP by taking the order w.r.t. the mark, see e.g.   M\o{}ller et al. (2016). 

Typically, a SSPP can be derived as the time-ordered vectors of points $x_i$. For a fixed $n$,  we denote by $\tt_n=(t_1, \dots, t_n)$ the ordered sequence of occurrence times $t_1 < t_2< \ldots < t_n$, and by  $\XX_n=(x_1, \ldots, x_n)$ the corresponding  ordered sequences of locations, while  we write $\left\lbrace x_1, \ldots, x_n \right\rbrace $ for the unordered sequence. The joint density of $(\XX_n, \tt_n)$ with respect to the Lebesgue measure on product space, can be presented naturally  as 
\begin{linenomath} 
	\begin{equation}\label{joint:density}
	f(\XX_n, \tt_n)= f_1(x_1, t_1) \prod_{k=2}^n f_k(x_k, t_k|\XX_{k-1}, \tt_{k-1})
	\end{equation}
\end{linenomath} 
where $f_1$ is the model for the time and location of the first point, and $f_k(x_k, t_k|\XX_{k-1}, \tt_{k-1})$ is the density for a new point $(x_k,t_k) $ conditional on the history of previous locations up to time $t_{k-1}$. The conditional probability densities in \eqref{joint:density} can be  decomposed into 
\begin{linenomath} 
	\begin{eqnarray}\label{decompos}
	f_k(x_k, t_k|\XX_{k-1}, \tt_{k-1}) &=& g_k(x_k |\XX_{k-1}, \tt_{k-1})
	\\ & & \cdot \,h_k(t_k|\XX_{k-1}, \tt_{k-1}, x_k)\nonumber
	\end{eqnarray}
\end{linenomath} 
where $g_k$ describes the spatial distribution of the $k$th event given the history up to time $t_{k-1}$, and $h_k$ represents the temporal distribution of the $k$th occurrence time given the history of the process up to time $t_{k-1}$ and the location of the $k$th point. Eventually, we note that the above decomposition agrees with the construction built in \citet{jensen2007} and in \citet{ylitalo2017} for the class of the STPP's. Now, since the structure of the SSPP lies merely on the spatial part in \eqref{decompos},  some assumption shall be imposed to allow us to recess the dominating role of time. We assume that the law of $x_k$ given the past does not depend on time. This yields
\begin{linenomath} 
	\begin{equation*}
	g_k(x_k |\XX_{k-1}, \tt_{k-1})=  g_k(x_k |\XX_{k-1}). 
	\end{equation*}
\end{linenomath}
%In fact, a stronger assumption could be introduced such as \textit{separability}

%and so
%\begin{eqnarray*}
%f(\XX_k, \tt_k) &= & g_1(x_1, t_1) \prod_{k=2}^n g_k(x_k|\XX_{k-1}) \\ & & \cdot \, h_1(t_1)  \prod_{k=2}^n h_k(t_k|\XX_{k-1}, \tt_{k-1}).
%\end{eqnarray*}
%
%One could also make the use of a stronger assumption such as \textit{separability}, i.e. space and time exert influences on the value of the process independently of each other. 
The SSPP model will be defined by this spatial component having the density, w.r.t. Lebesgue measure, of the form of
\begin{linenomath} 
	\begin{equation}\label{lh}
	g(\XX_n)= g_1(x_1) \prod_{k=1}^{n-1} g_{k+1}(x_{k+1}|\XX_k),
	\end{equation}
\end{linenomath}
as a marginal distribution extracted from the STPP model. Yet, the history information drawn by  the spatial locations is preserved under the successive conditioning.
%In general, other marginal distributions might be of interest such as the ordered sequence of times $\{ \tt_n\}$ on $T$, 

%For our observed sequence $\XX_n=(x_1, \ldots, x_n)$, we are interested in the sequential point process modeled by 
%\begin{equation*}\
%f(\XX_n)= f_1(x_1) \prod_{k=1}^{n-1} f_{k+1}(x_{k+1}|\XX_k)
%\end{equation*}

%\texttt{*** marginal distribution}

\subsection{The Parametric SSPP Model}

%the likelihood of a sequential point process model
%can be written by using conditional densities.

In a bounded window $W \subset \R^2$, let the observed data $(x_i, m_i),  i=1, \ldots, n$, with finite $ n \in \N_0 $, be a set of points $x_i$ indicating the locations of trees, and a set of corresponding marks $m_i \in \R$ representing the sizes of the trees which are the DBH's, and such that the following order  $0< m_1 < m_2< \ldots <m_n < \infty $ holds, where $m_1$ and $m_n$ are the sizes of the smallest and the largest tree respectively. 

The data can be modeled by a STPP model, or in a quite natural way, by a marked point process (MPP) model.
% where the marks are taken as the  event times $t_i=m_i$. In contrast, one could consider a marked point process (MPP) instead.
However, there is  a one-to-one correspondence between the class of the MPP's and the class of the STPP's such that a MPP can be transformed into a STPP by taking the order with respect to the marks \citep{moller2016}, or conversely by treating all event times $t_i$ as marks \citep{daleyverejones2008,verejones2009,stoyan2017}.  In this paper, modeling the time events will not play any importance since we are concerned only with the spatial dimension.
%%We refer to this  ordered sequence of marks by $\mmm=(m_1, \ldots, m_2) $.
%% As one could arranges the marks 
%The data can be viewed as a realization of a STPP where the occurrence times are $t_i=m_i$. In fact, one could consider a marked point process instead

From the observed data we write the  ordered sequence of points $\XX_n=(x_1, \ldots, x_n)$ with respect to the order of the $m_i$'s. Following Section  \ref{prelim}, the density function of a SSPP process is given by \eqref{lh}.

In order to give the expression \eqref{lh} an explicit form, two main principles are considered: (1) the model should catch the long-term dependence between the locations $x_i$'s and read the memory of the underlying dynamics, and  (2) the parameters of the model interpret these effects into a  self-interaction framework.
%can be interpreted in terms of inter-tree competitionwe express the SSPP model

For a real number $r>0$ and $ k\in \{2, \dots,n-1\}$, we define the \textit{lagged clustering measure} $\Sr(\XX_k, y)$ which counts the number of earlier balls $B(x_i, r), i =1, \ldots,k,$ that contains the new point $y$:
\begin{linenomath} 
	\begin{equation}\label{rec}
	\Sr(\XX_k, y)= \sum_{i=1}^{k}\1_{B(x_i,r)}(y).
	\end{equation}
\end{linenomath} 
Based on this measure, we introduce the self-interaction function $\pp( y, \Sr(\XX_k, y))$  as a reweighting  probability of the forthcoming point $y$ in the form of
\begin{linenomath} 
	\begin{equation}\label{selinteractionmeas}
	\pp( y, \Sr(\XX_k,y))=\begin{cases}
	\theta & \text{if $\Sr(\XX_k, y) \geq 1$},\\
	1-\theta & \text{if $\Sr(\XX_k, y) =0$},
	\end{cases}
	\end{equation}
\end{linenomath} 
%	or under the compact form of
%\begin{eqnarray*}
%\pp( x_{k+1}, \Sr(\XX_k, x_{k+1})) &=& \theta \, \1_{\{\Sr(\XX_k, x_{k+1}) \geq 1\}}(x_{k+1})\\ & & + (1-\theta) \,  \1_{\{\Sr(\XX_k, x_{k+1})=0 \}}(x_{k+1}).
%\end{eqnarray*}
where $\theta \in (0,1)$. Now, we define the SSPP model at the  new location $y$,  conditioning on the sequence $\XX_k$,  by taking the conditional density function 
\begin{linenomath} 
	\begin{equation}\label{fullmodel}
	g_{k+1}(y|\XX_k) \wasypropto  \pp( y, \Sr(\XX_k, y)),
	\end{equation}
\end{linenomath} 
or with the compact form of
\begin{eqnarray*}
	g_{k+1}(y|\XX_k) &\wasypropto& \theta \, \1_{\{\Sr(\XX_k, y) \geq 1\}}(y)\\ & & + (1-\theta) \,  \1_{\{\Sr(\XX_k, y)=0 \}}(y).
\end{eqnarray*}
The  self-interaction function \eqref{selinteractionmeas}  first appeared in the work of \citet{penttinen2016} in modeling eye movement, and under \eqref{fullmodel}, the parametric SSPP model is considered as a special case of the model used by \cite{penttinen2016} defining a \textit{history-dependent} model with parameters  $\theta$ and  $ r$. Apparently, our model is using the full past history to read the long-term dependency of $y$ on the $\XX_k$. However, other ranges of history can be  also of  interest, such as restricting the history up to  the last $m$ points $x_{k-m}, \ldots, x_{k-1}$, $m <k$, which is defining the so-called $m$-memory point processes where  $m=1$ stands for the Markovian case, see \citet{synder}. The authors intend to investigate these types of history under the sequential approach in future work.

The component $\Sr(\XX_k, y)$ is thought to examine the  long-term spatial dependence between the new location $y$ and the past locations  $x_i \in \XX_k$ by interpreting this evidence into the parameter $\theta$, which is in fact  
%by evaluating the number of  points in $\XX_k$ which lie closer to $y$ than the distance $r$.
%The parameter $\theta$
the probability that the point $y$ lies inside, at least, in  one of the balls $B(x_i, r), i =1, \ldots,k,$ formed by the past $\XX_k$. While the probability $1-\theta$ is for the case when $y$ lies outside of all the balls $B(x_i,r), i =1, \ldots, k$. 

%    From \eqref{rec} and \eqref{fullmodel}, the approach of the self-interaction  introduced by  the SSPP  model  is useful in  the sense that it emphasizes the attraction or the repulsiveness depending on the value of $\theta$ inside this interaction area.   In particular, if  the parameter $\theta$ is close to $1$ the model  accepts the birth of the new locations in the  neighborhood $B(x_i, r)$ of the previous points with a higher probability, which leads to clustering. On the other hand, a lower value of $\theta$ indicates that the model favors new locations in the non-visited areas rather than the locations in the neighborhood areas of  the previous points. In the case of $\theta=0.5$, the model is identified as a random walk without self-interaction.  

Endowed with  the parameters $\theta$ and $r$,  the model is allowed to perceive  spatial features of the data such as the interaction area of a given point $x_i, i =1, \ldots,k$, which is defined by the ball  $B(x_i,r)$,  and the spatial coverage of the sequence over the window $W$ which is the union of all the balls  $B(x_i,r), i =1, \ldots,k$. These features serve as tools to build the  summary statistics when comparing the data to the simulated realizations of the fitted model. From \eqref{rec} and \eqref{fullmodel}, the approach proposed by the self-interaction    is useful in  the sense that it emphasizes the attraction or the inhibition depending on the value of $\theta$.   In particular, if  the parameter $\theta$ is close to $1$ the model  accepts the birth of the new locations in the  neighborhood $B(x_i, r)$ of the previous points with a higher probability, which leads to clustering. On the other hand, a lower value of $\theta$ indicates that the model favors new locations in the non-visited areas rather than the locations in the neighborhood areas of  the previous points. In the case of $\theta=0.5$, the model is identified as a random walk without self-interactions, and  consequently  the density \eqref{lh} is independent of the order in this case (see Section 5).

\section{Statistical Inference}
\label{s:inf}

\subsection{Likelihood}
We estimate the parameters involved in the model by using the maximum likelihood (ML) method.   This method has been applied for finite sequential point processes such as random sequential absorption model \citep{lieshout2006c}, and sequential model for eye movement \citep{penttinen2016}. The likelihood function for our SSPP model can be expressed up to a normalizing constant as
% . For instance, the likelihood for the general sequential spatial point process model with all three
%components introduced above can be presented in form
%Now, Given a SSPP with  density  \eqref{lh} and self-interaction function \eqref{fullmodel} with unknown parameters $\theta$ and $r$, an  expression of the  likelihood for the  SSPP in terms of parameters can be presented in form of
\begin{linenomath} 
	\begin{equation}\label{LH}
	\LL(\XX_k)=  g_1(x_1) \prod_{k=1}^{n-1} \alpha_k^{-1} \,\pp( x_{k+1}, \Sr(\XX_k, x_{k+1})),
	\end{equation}
\end{linenomath} 
where $\alpha_k=\int\limits_{W}  \pp( u, \Sr(\XX_k, u))\, \dd u $ is the normalizing constant depending on the window $W$, and on the parameters $\theta$ and $r$. In fact, the normalizing constant is analytically intractable, but the existing numerical integration methods can be used to compute the integral.
% We have then 
%	$$
%\LL(\XX_k)=  f_1(x_1) \prod_{k=1}^{n-1} \frac{\pp( x_{k+1}, \Sr(\XX_k, x_{k+1}))}{\int\limits_{W}  \pp( u, \Sr(\XX_k, u))\, \dd u}
%$$
From \eqref{LH}, the likelihood  is integrable since it is bounded, which makes the model well-defined. We write then  the  log-likelihood for the model in the form:
\begin{eqnarray}\label{loglh}
l(\theta, r)&=& 
\log(\theta) \sum_{k=1}^{n-1} \1_{\{\Sr(\XX_k, x_{k+1}) \geq 1\}}(x_{k+1}) \nonumber  \\
& &+\log((1-\theta)) \sum_{k=1}^{n-1} \1_{\{\Sr(\XX_k, x_{k+1}) =0\}}(x_{k+1}) \nonumber \\
& & - \sum_{k=1}^{n-1} \log \int\displaylimits_{W} \theta \, \1_{\{\Sr(\XX_k, u) \geq 1\}}(u) \nonumber\\& & + (1-\theta) \, \1_{\{\Sr(\XX_k, u) =0\}}(u)\,   \dd u.
\end{eqnarray}
Due to the normalizing integral, maximizing the log-likelihood can have computational burden. Here we choose to evaluate the log-likelihood over a grid of values of pairs $(\theta, r) \in (0,1) \times (\underline{r}, \overline{r})$ in order to find the maximum. Alternatively, similar numerical optimization methods can be used such as the profile likelihood approach, see \citet{moller2004} and \citet{davison2008}. The values $\underline{r}$ and $ \overline{r}$ are the lower/upper bounds for the parameter $r$ chosen according to the given data; we shall refer to these bounds in Section 4. 

\subsection{Model Evaluation}\label{summary}
%Fitting the model  should be accompanied by  measures assessing the uncertainty. knowledge about the statistical variation
%
Assessing the goodness-of-fit of the model will be conducted by using the envelope method in order to indicate the statistical variation in the summary statistic under the parametric  model assumption. The envelope method
%first introduced by Ripley (1977).
allows to compare the empirical functional summary statistics estimated from the data with the same summary statistic obtained through simulations of the fitted model based on  the parametric bootstrap approach \citep{efron1994}.  

In our case, we need to employ metrics that measure different characteristics of the data under the sequential approach. In contrast, the usual summary functions used in the field of spatial statistics lack the ordering feature. Therefore,  in order to  describe the dynamics of the SSPP model,  we  establish various functional summary statistics as functions of time, and also that are justified from the forest applications point of view to measure competition of trees for the resources (light, water, nutrients). Accordingly, we  utilize four functional summary statistics that are related to  self-interaction features and to  the area coverage of the sequence \citep[see][]{penttinen2016}, and also summaries related to the nearest-neighborhood distances and  the area interaction.
%These statistics are all given in the cumulative form. 

% << echo=FALSE, fig=TRUE, message=F, warning=F >>=
% library(spatstat)
% load('.RData')
% plot(AIdemo, main="")
% lines(x,y, lwd=2)
% hpts <- chull(x,y)
% hpts <- c(hpts, hpts[1])
% lines(x[hpts], y[hpts], lwd=1, lty=3, col = "magenta")
% @

\begin{enumerate}[wide]
	\item {\itshape Lagged clustering statistic:} It is based on the lagged clustering measure $\Sr(\XX_k, x_{k+1})$ which calculates the number of earlier points $x_i$'s in $\XX_k=(x_1, \dots, x_k)$ closer  to the new point $x_{k+1}$ within the range of the radius $r$. The variation along the time (ordered points) axis of this measure shall indicate how frequently the birth of new trees is close spatially to the earlier	trees.	
	
	%For this statistic we use the cumulative version of the measure.
	
	\item {\itshape  First contact distance:} We consider the distance between a given new point $x_{k+1}$ and the past sequence $\XX_k$. We define it as the minimum of all
	the distance $\left\| x_{k+1} - x_i \right\|$, where $\left\| \cdot \right\| $ denotes the Euclidean distance and $x_i \in \XX_k$, i.e. $\min_{i\leq k} \left\| x_{k+1} - x_i \right\|$. This measure acts as the sequential counterpart of the nearest-neighbor	distance or the void distance, but with respect to the past only. 
	
	%\item {\itshape Area-interaction statistic:} When a new point is interacting with the past points, we need to measure the strength (or weakness) of this interaction. Considering the ball $B(x_{k+1}, r)$ centered at location $x_{k+1}$ and with a radius $r$, we examine the part of this region that is influenced by  $ \bigcup_{k=1}^k B(x_i, r)$ formed by past locations $x_i, i=1 ,\ldots, k$. This overlapping region will determine the zone for $x_{k+1}$ that is  reached by  other past $x_i, i=1 ,\ldots, k$, which  tell how much the new tree $k+1$ competes for resources when approaching the
	%older trees. We quantify this zone by computing its normalized area as
	%$$
	%c_i= \frac{\left|  B(x_i,r)\,\bigcap \left\lbrace \bigcup\limits_{k=1}^{i-1}  B(x_k,r) \right\rbrace \cap W  \right| }{\left| B(x_i,r)\cap W\right| }
	%$$
	%where the symbol $\left| \cdot\right|$ denotes the area.  This statistic  is useful for showing local characterization of the interaction at each point. 
	
	\item {\itshape Proper  zone statistic:} When a new point is interacting with the past points, we postulate a characterization of this interaction   \textit{locally} in order to see  how strong, or weak, is the connection of the new location to the past ones  in terms of the neighborhood region of the new location. Considering the ball $B(x_{k+1}, r)$ centered at location $x_{k+1}$ and with a radius $r$, we examine the part of this region that does not intersect with any other balls $B(x_i, r)$ formed by past locations $x_i, i=1 ,\ldots, k$. This will determine a {\em proper zone} for $x_i$ that is not reached by any other competitor, which is in fact the complementary of  the overlapping region  $ B(x_i,r)\,\bigcap \left\lbrace \bigcup\limits_{k=1}^{i-1}  B(x_k,r) \right\rbrace$  inside the ball $B(x_{k+1},r)$.  We quantify the proper zone statistic by measuring its normalized area with the value:
	\begin{linenomath} 
		$$
		c_{k+1}= \frac{\left| \left\lbrace B(x_{k+1},r)\, \backslash \bigcup\limits_{i=1}^{k}  B(x_i,r) \right\rbrace \cap W  \right| }{\left| B(x_{k+1},r)\cap W\right| }
		$$
	\end{linenomath} 
	where the symbol $\left| \cdot\right|$ denotes the area. In the case where the points $x_i$ represent the locations of trees with neighborhoods $B(x_i,r)$, the proper zone statistic induces the resources territory  used by a tree without being in a competition with larger trees, while the overlapping zone indicates in this case the space where the tree competes for resources in its neighborhood $B(x_{k+1}, r)$.
	
	\item {\itshape Ball union coverage:} At each point $x_k$, we consider the \textit{regionalized} form of the sequence $\XX_k$ by taking  the union  $\bigcup_{i=1}^k B(x_i,r) \cap W$. This shall   measure the coverage and the degree of filling cast by the sequence at each time in $W$, which  also  indicates the development of empty space inside $W$. We write it in a  normalized form as 
	\begin{linenomath}  $$\frac{\left| \bigcup\limits_{i=1}^k B(x_i,r) \cap W \right| }{\left| W \right| }.$$ \end{linenomath}

\end{enumerate}

\section{Simulation study}
%Monte-Carlo simulations are often a single reliable solution method in many
%financial problems. Within the simulation study the random variables are generated from some prescribed distributions
%
In this section we examine the behavior of the SSPP model via simulation, where we generate realizations  of the suggested model  using conditional distribution \eqref{fullmodel} with parameters $(\theta, r)$. Simulated realizations are drawn from  different classes of the SSPP model depending on the self-interaction character. Thus, we will employ the summary statistics proposed above in order to see how well the statistics could capture the features of  simulated data. 

Each realization consists of a sequence $\XX_{100}$ consisting of 100 points located in the unit square window, and  for the sake of simplicity,   the first two points $x_1$ and $x_2$ of the sequence $\XX_{100}$ shall be drawn uniformly and will be used as starting points  for all realizations. The first sampled points are $x_1 = (0.90,0.50)$ and $x_2=(0.60, 0.92)$. Later, the sequence is simulated by adding points $x_k$ using the accept-reject method \citep{ripley1987} such that the law of $x_k$ given the past $\XX_{k-1}$ follows the density \eqref{fullmodel}. 

Our interest in simulation is to observe the variability of the self-interaction function among three different types of SSPP models. To this end, a certain range of $r$ has to be given, which will be fixed at  $r=0.1$. With respect to  the parameter $\theta$, we propose  the following models:  \\
- Model 1:  $\theta=0.05.$\\
- Model 2:  $\theta =0.5.$\\
- Model 3:  $\theta= 0.95.$

\begin{figure}
	\begin{center}
		%	LAST USED	\centerline{\includegraphics[width=9cm]{simp.pdf}}
		\centerline{\includegraphics[width=9cm]{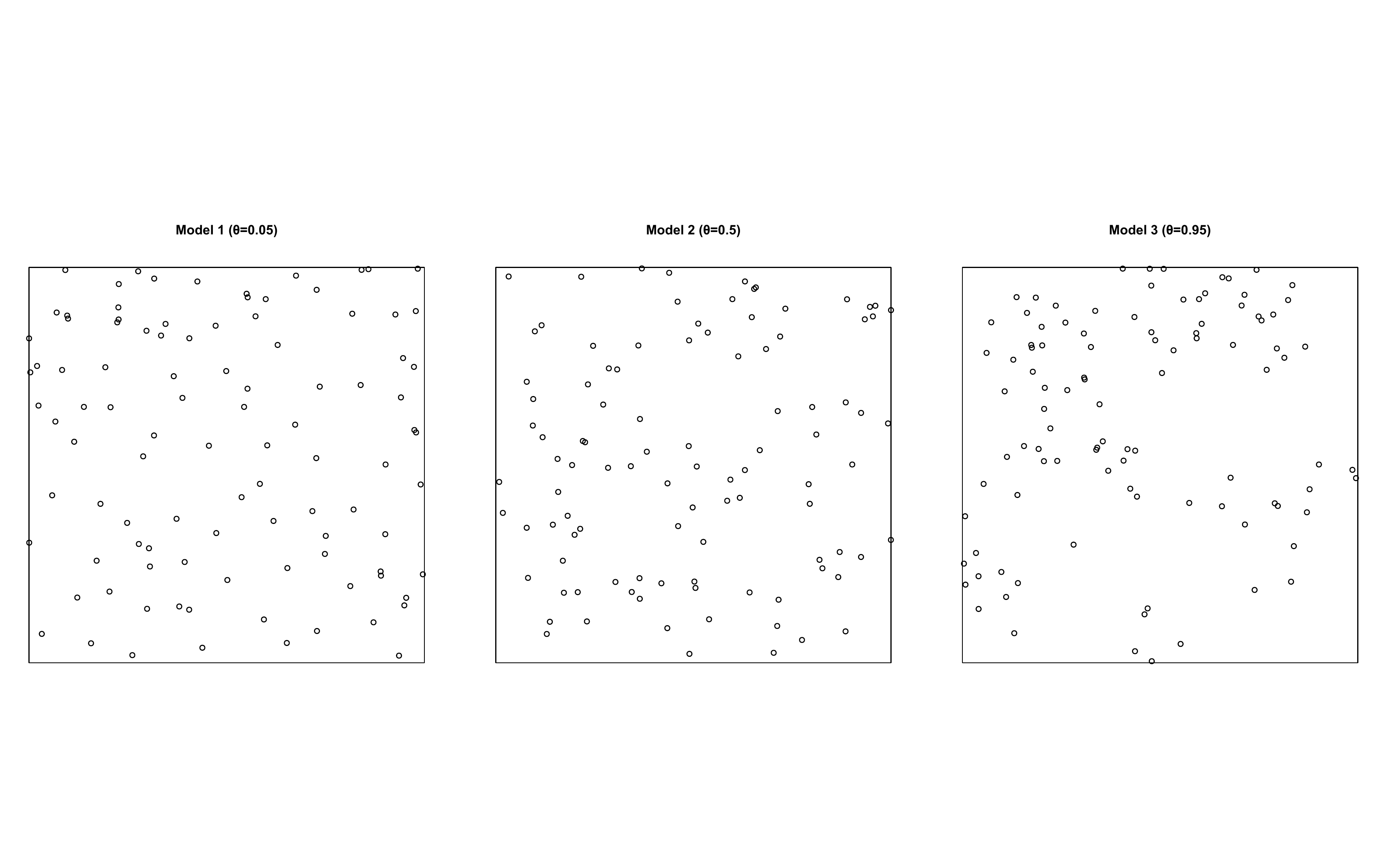}}
	\end{center}
	\vspace*{-1.cm}
	%\begin{center}
	%\centerline{\includegraphics[width=9cm]{simL.pdf}}
	%	\end{center}
	\caption{ \small Simulated realizations of the three models: Model 1 (left), Model 2 (middle) and Model 3 (right).}
	\label{sim:pattern}
\end{figure}

\begin{figure}
	\begin{center}
		%		\centerline{\includegraphics[width=8cm]{simualtionlast.pdf}}
		%\centerline{\includegraphics[width=8cm]{simulation2.pdf}}
		%\centerline{\includegraphics[width=8cm, scale=0.4]{simulation3.pdf}}
		%\centerline{\includegraphics[width=8cm]{simulation5.pdf}}
		\centerline{\includegraphics[width=8cm]{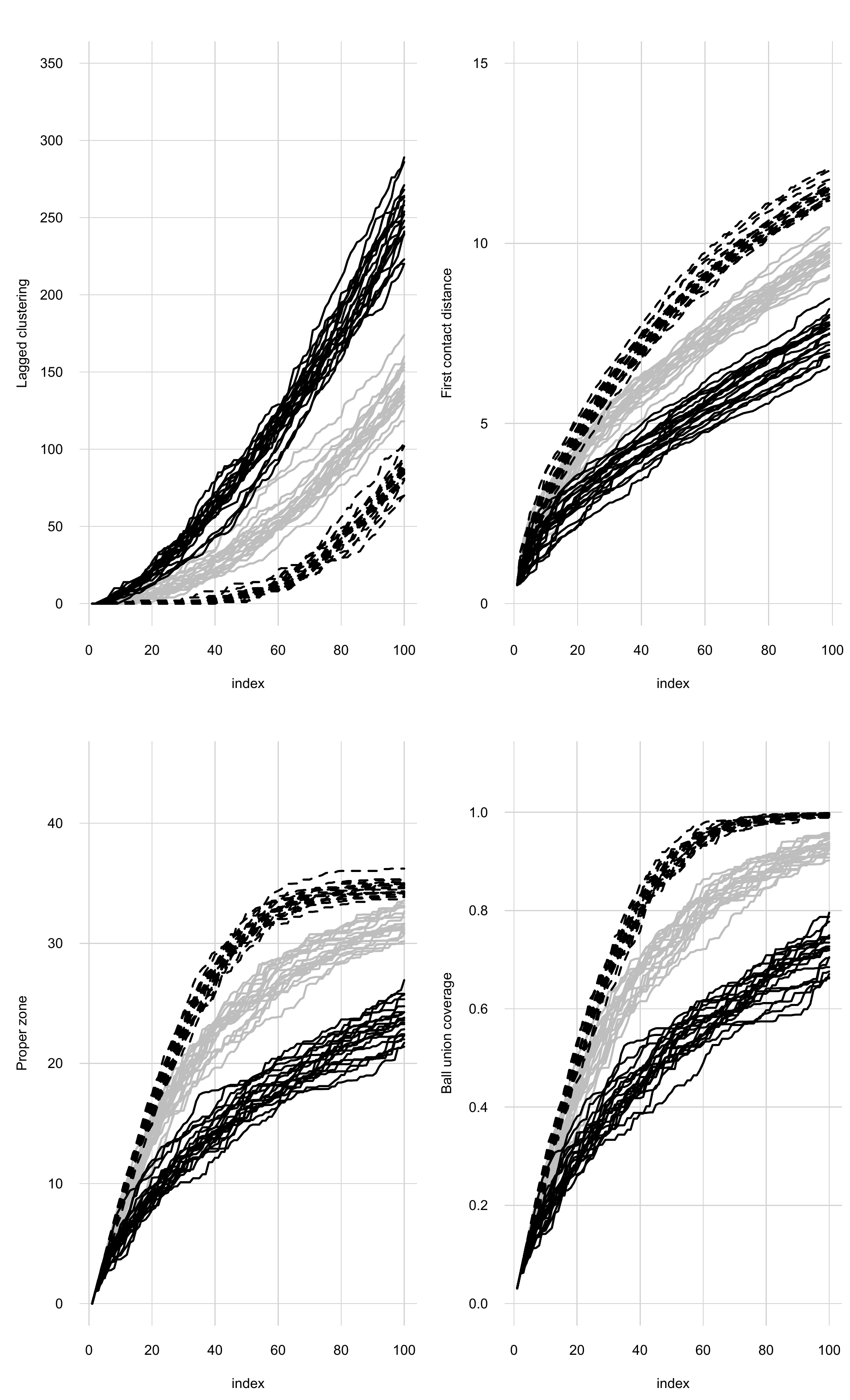}}
	\end{center}
	%\vspace*{-0.915cm}
	%\begin{center}
	%\centerline{\includegraphics[width=9cm]{simp.pdf}}
	%	\end{center}
	\caption{ \small Summary statistics of simulated patterns of three different SSPP models. Cumulative lagged clustering statistic (top left), cumulative first contact distance (top right), cumulative proper zone statistic (bottom left), and the cumulative ball union coverage (bottom right). Dark solid lines are statistics of the Model 3, dashed lines represent those of Model 1, and grey lines are the statistics of the Model 2.}
	\label{sim:fig}
\end{figure}

We generate 20 realizations of each model and we compute the four functional summary statistics. A realization of simulated patterns of each model is visualized in Figure \ref{sim:pattern}.  In  Figure \ref{sim:fig}, the results related to the summary statistics are given in the cumulative form. The lagged clustering statistic based on the measure \eqref{rec} computes the number of earlier points near the current point, say $x_k$, and the cumulative version sums all these numbers together, i.e. $\sum_{i=2}^k \Sr(\XX_{i-1}, x_i)$. We see that Model 1 avoids the locations nearby
other points when compared to the random walk, and its related cumulative first contact distance statistic is taking larger values expressing this avoidance character of the pattern compared to  Model 3. This latter  exhibits high values of cumulative lagged clustering statistic where the points are favoring locations nearby previous ones and the first contact distances are the smallest among these three models, indicating a tendency towards higher clustering.

The ball union coverage reveals that the coverage of the Model 1 increases faster than for the other two models, filling almost the whole window just after 60 points, and the forthcoming points indexed from 60 to 100 try to locate themselves nearby the former ones, which is  indicated by  a slow-up of the ball union coverage function. Accordingly, the fast coverage of Model 1 is consistent with the increase of the proper zone gained by the points as can be seen from the proper zone statistic. On the other hand, the coverage of Model 3 fills only less than 80\% of the window expressing the clustering effect gathering the new points near the former ones. In addition, the proper zone statistic of Model 3 exposes a similar effect.

%\begin{figure}
%	\begin{center}
%	\begin{tikzpicture}
%	\node[anchor=north west ,inner sep=0] (frame1) at (0,0)        
%	{\includegraphics[width=0.45\textwidth]{rec_simulated.pdf}};
%	
%	\node[anchor=north east,inner sep=0] (frame2) at (0,0)        
%	{\includegraphics[width=0.45\textwidth]{area_simulated.pdf}};
%	
%	\node[anchor=south west,inner sep=0] (frame3) at (0,0)        
%	{\includegraphics[width=0.45\textwidth]{path_simulated.pdf}};
%	
%	\node[anchor=south east,inner sep=0] (frame4) at (0,0)        
%	{\includegraphics[width=0.45\textwidth]{hull_simulated.pdf}};
%	\end{tikzpicture}
%	\caption{My caption}
%\end{center}
%\end{figure}

\section{Application}
\label{app}

%Please see the file \texttt{biomsample.tex} for fancy examples of making
%tables.  Here is a very simple one.  Use \texttt{table} for tables
%that are narrow enough to fit in one column of the typeset journal; use
%\texttt{table*} for tables that need to span two columns.  For
%figures, use of \texttt{figure} and \texttt{figure*} is analogous. 
%
%\begin{table}
%\caption{This is a simple table.}
%\label{t:one}
%\begin{center}
%\begin{tabular}{lrrr}
%\Hline
%Estimator & \multicolumn{1}{c}{$\beta_1$} &  \multicolumn{1}{c}{$\beta_2$} & 
%\multicolumn{1}{c}{$\beta_3$} \\ \hline
%MLE & 10.18 & $-$3.26 & 0.13 \\
%OLS & 9.92 & $-$3.19 & 0.11 \\
%WLS & 9.88 & $-$3.33 & 0.12 \\
%\hline
%\end{tabular}
%\end{center}
%\end{table}
%
%You can experiment with fancier tables than Table~\ref{t:one}.
%
%We can get bold symbols using \verb+\bmath+, for example, $\bmath{\alpha}_i$.

\begin{figure}
	\begin{center}
		\centerline{\includegraphics[width=10.5cm]{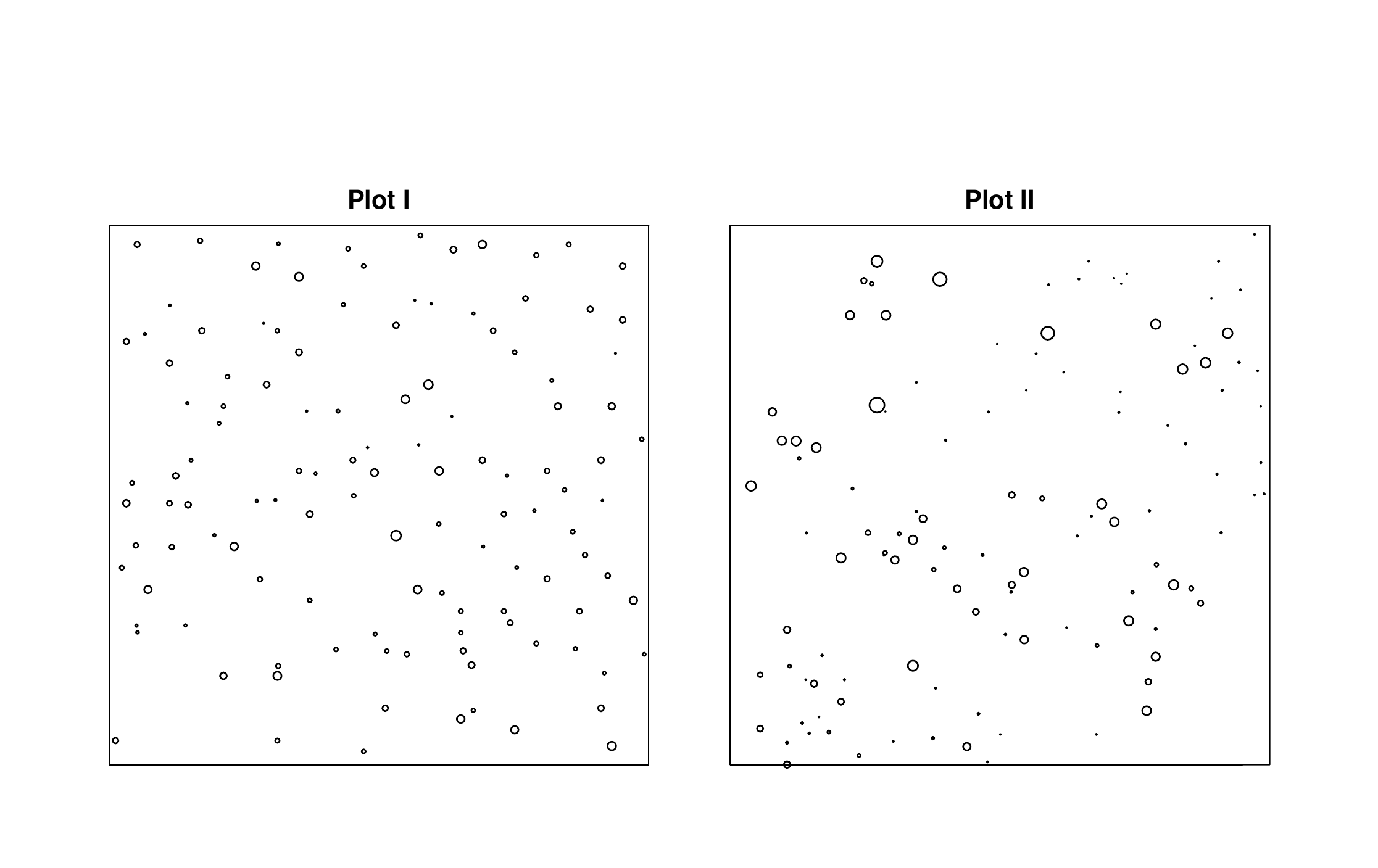}}
		%	\centerline{\includegraphics[width=8.5cm, height=8cm]{5_74_L.pdf}}
		\centerline{\includegraphics[width=8.5cm, height=8cm]{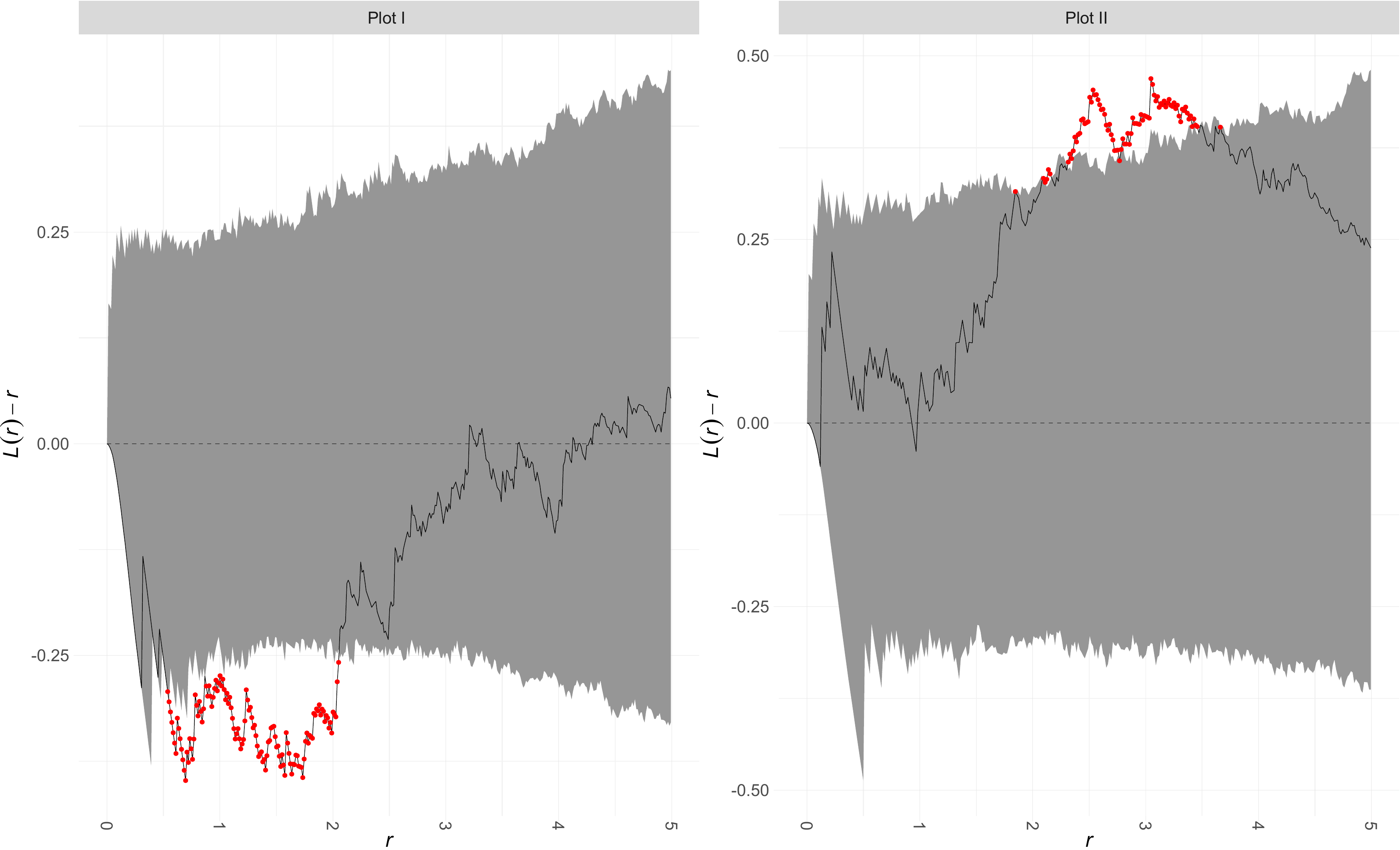}}
	\end{center}
	\caption{\small First row: Positions of $120$  trees (Plot I) and of $118$ trees (Plot II) in a $25 \times 25$ meters 	sampling region in Kiihtelysvaara  (Eastern Finland), where the radius of a disc is $2 \times$DBH. Second row: Global envelope test for the summary $L(r)-r$  for Plot I (left) and Plot II (right) based on 4999 simulation of CSR. The grey areas show the 95\% global envelopes on the interval [0,5]. The solid black line is the data function and the dashed line represents the (estimated) theoretical expectation.
		\label{plot}}
\end{figure}

To demonstrate the performance of the  SSPP model constructed in Section 2, we need to apply it to forest stand data. Our data  was collected from Kiihtelysvaara site situated in  Northern Carelia, Eastern Finland (more details on stand data can be found in \citealt{packalen2013}). The dataset includes  79 forest plots, however,  in order to  illustrate the machinery of the model,  we work with  two  sample plots exhibiting two different spatial structures, namely Plot I and Plot II. A summary of estimated parameters of the model for all the 79 plots   is given in  Appendix.   Dataset    consists  of tree locations and stem diameters measured at breast height. The Plot I contains 120 trees with minimum and maximum DBH of 4.10 cm and 23.15 cm   having the mean value of 11.39 cm. In Plot II, there are 118 trees with DBH ranging from 2.30 cm as a minimum value to 41.95 cm as a maximum value, and with the mean of 12.11 cm.  Figure \ref{plot} (first row) shows the patterns describing the spatial structure in  Plot I and  Plot II in bounded windows  of sizes $25 \times 25 \,\textrm{m}^2 $.   In our analysis, we ignore any edge effects and all the variation of the model shall be restricted to the bounded window.

A preliminary stage of the study would be an exploratory analysis using the second-order structures of the \textit{static} point patterns presented in Figure \ref{plot} (first row). At this point, we adopt  the linearized version of Ripley’s K-function \citep{ripley1977,lotwick1982} which is based on the $\K$-function via the formula $\LL(r) =\sqrt{\K(r)/ \pi}$, where $r$ is a distance between a pair of points. More  details on $\LL$-function can be found in \citet{illian2008}. Taking the centered form $\LL(r)- r$ of the $\LL$-function at a corresponding scale, the complete spatial randomness (CSR) is characterized by  $\LL(r)- r = 0$, while  values   of $\LL(r)- r$ larger than $0$ indicate clustering; the case where $\LL(r)- r <0 $    expresses regularity. 

To test the complete spatial randomness hypothesis, a global envelope testing  is carried out using the extreme rank length (ERL)  ordering \citep{mylly2017,mrv2018}. The test is conducted using 4999 simulations of CSR on  chosen interval [0,5] (in meters) for $r$, and the evaluation  is represented graphically in Figure \ref{plot} (second row) indicating a rejection of the CSR hypothesis for both of the plots. The corresponding p-values of the tests are reported as $p<0.001$ and $p=0.014$ for Plot I and Plot II respectively. Apparently, the centered $\LL$-function estimated from Plot I had negative values for  distances 0.53-2.05 m  below the lower envelope, which means that spatial structure of Plot I is regular at that scale. In contrast, the estimated centered $\LL$-function of Plot II shows a clustering character at scale 1.80-3.66 m.

%the CSR as null model is usually simulated $n$ times and the estimate of $L_{min}(r)$ and $L_{max}(r)$  is calculated for each $r$, so that  if  $\hat{L}(r)$, the estimate for data, lies outside the envelope bounded by  $L_{min}(r)$ and $L_{max}(r)$, the null hypothesis is rejected at that range $r$. Critisim on this pointwise envelope testing was discussed by 
%
% a standard approach carried out based on global envelope
%
%The analysis of patterns of trees different levels revealed that between-levels interactions have a character

From the observed locations of trees in each plot, we form an ordered sequence $\XX_n=(x_1, \ldots, x_n)$ of these locations in a decreasing order based on DBH, where $x_1$ is the location of the largest tree and $x_n$ the location of the smallest tree in the plot. 
%This order seems natural from ecological point of view since the largest trees appear first, however, we exclude here the cases where a trees outside the window might be naturally a member of the sequence, so our model seems to be ideal in this sense, and so,  
%we ignore  edge effects and all the variation of the model shall be restricted to the bounded window.

To estimate the model parameters $\theta$ and $r$ we need to set up the ranges of these parameters. For specificity, the range of $\theta$ is obviously the interval $(0,1)$, while for the parameter $r$  we take $(\underline{r}, \overline{r})$ to be the range of competition strength $(0,5)$. However, values of $r$ near the origin zero may be problematic because the stems cannot overlap. Therefore,  since  we lack information on ground measurements and biological knowledge, we take $\underline{r}$ to be the radius of  the largest tree in the plot, so we write $\underline{r}= 0.11575$ meters for Plot I and $\underline{r}=0.20975$ meters  for Plot II.

Computing the integral in \eqref{loglh} requires numerical integration tools, hence, we used  the Riemann sums method. To find  the maximum likelihood estimate, the function is maximized over a grid of values of $(\theta,r)$ with steps $0.05$ and $0.1$ for $\theta$ and $r$ respectively. In addition, we rechecked the results by using the optimization routines implemented in the R package \verb|nloptr| \citep{johnson2010}. Firstly, we employ the deterministic-search algorithm  \verb|Direct-L|   \citep[see][]{gablonsky2001} to obtain the global optimum and secondly we adjust it for greater accuracy using the  Nelder-Mead simplex algorithm \citep{richardson1973}.  The maximum likelihood estimates for the Plot I and Plot II are summarized in  Table \ref{t:one}, together with the 95\% bootstrap confidence intervals calculated from 20 realizations of the fitted model \citep[see][]{efron1994}.

\begin{table}
	\caption{\small Parameter estimates of SSPP model for Plot I and Plot II. The first and third columns  contain MLEs obtained for the data. The second and fourth columns  contain confidence intervals for the estimated parameters based on 95\% bootstrap confidence intervals.}
	\label{t:one}
	\begin{center}
		\begin{tabular}{lrrrr}
		%	\hline
			& \multicolumn{1}{c}{$\hat{\theta}$} &  \multicolumn{1}{c}{$95$\% CI for $\hat{\theta}$ } & \multicolumn{1}{c}{$\hat{r}$} & 
			\multicolumn{1}{c}{$95$\% CI for $\hat{r}$ }  \\ \hline
			Plot I & 0.17 & (0.13, 0.20) & 2.18 & (2.11, 2.25)\\
			Plot II & 0.65 & (0.58, 0.74) & 2.60 & (1.06, 4.13)\\
			\hline
		\end{tabular}
	\end{center}
\end{table}

Next we evaluate the SSPP model by computing the four summary statistics, mentioned earlier in Section \ref{summary}, from the data and from 999 simulated realizations of the fitted model. When simulating the model, we use the observed  values of the first two locations  as starting points of the simulated sequence in order to  reduce the unexpected variations of the simulated realizations. 
% \begin{figure}
%	\begin{center}
%		\centerline{\includegraphics[width=8.5cm]{test5.pdf}}
%	\end{center}
%	\caption{cumulative recurrence function (top left), convex hull coverage (top right), path length of the sequence(bottom left) and convex hull coverage area (bottom right) for Plot I (black solid line). Grey area  represents pointwise envelopes estimated from 999 simulations of fitted model. \label{envplot1}}
%\end{figure}
%
%  \begin{figure}
%	\begin{center}
%		\centerline{\includegraphics[width=8.5cm]{test74.pdf}}
%	\end{center}
%	\caption{cumulative recurrence function (top left), convex hull coverage (top right),
%		path length of the sequence(bottom left) and convex hull coverage area (bottom right)
%		for Plot II (black solid line). Grey area  represents pointwise envelopes estimated from 999 simulations of fitted model.\label{envplot2} }
%\end{figure}

\begin{figure}
	\begin{center}
		%   \centerline{\includegraphics[width=8cm]{5_74.pdf}}	
		%	\centerline{\includegraphics[width=8cm]{5_74_largebox.pdf}}
		% WAS THE LAST USED\centerline{\includegraphics[width=8cm]{5_74_1sum.pdf}}
		\centerline{\includegraphics[width=8cm, height=9cm]{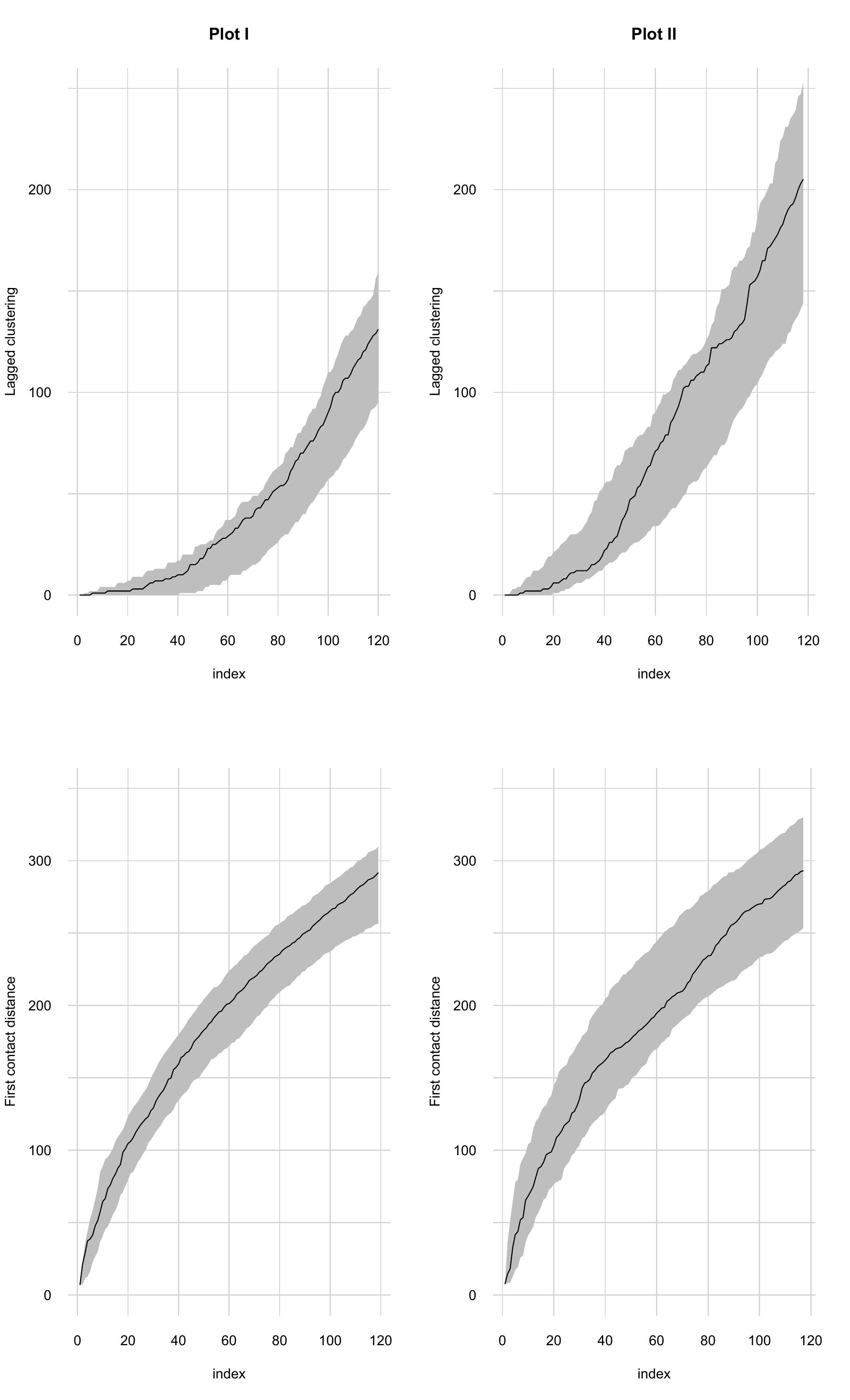}}
		% was the alst used \centerline{\includegraphics[width=8cm]{5_74_2sum2.pdf}}
		\centerline{\includegraphics[width=8cm, height=9cm]{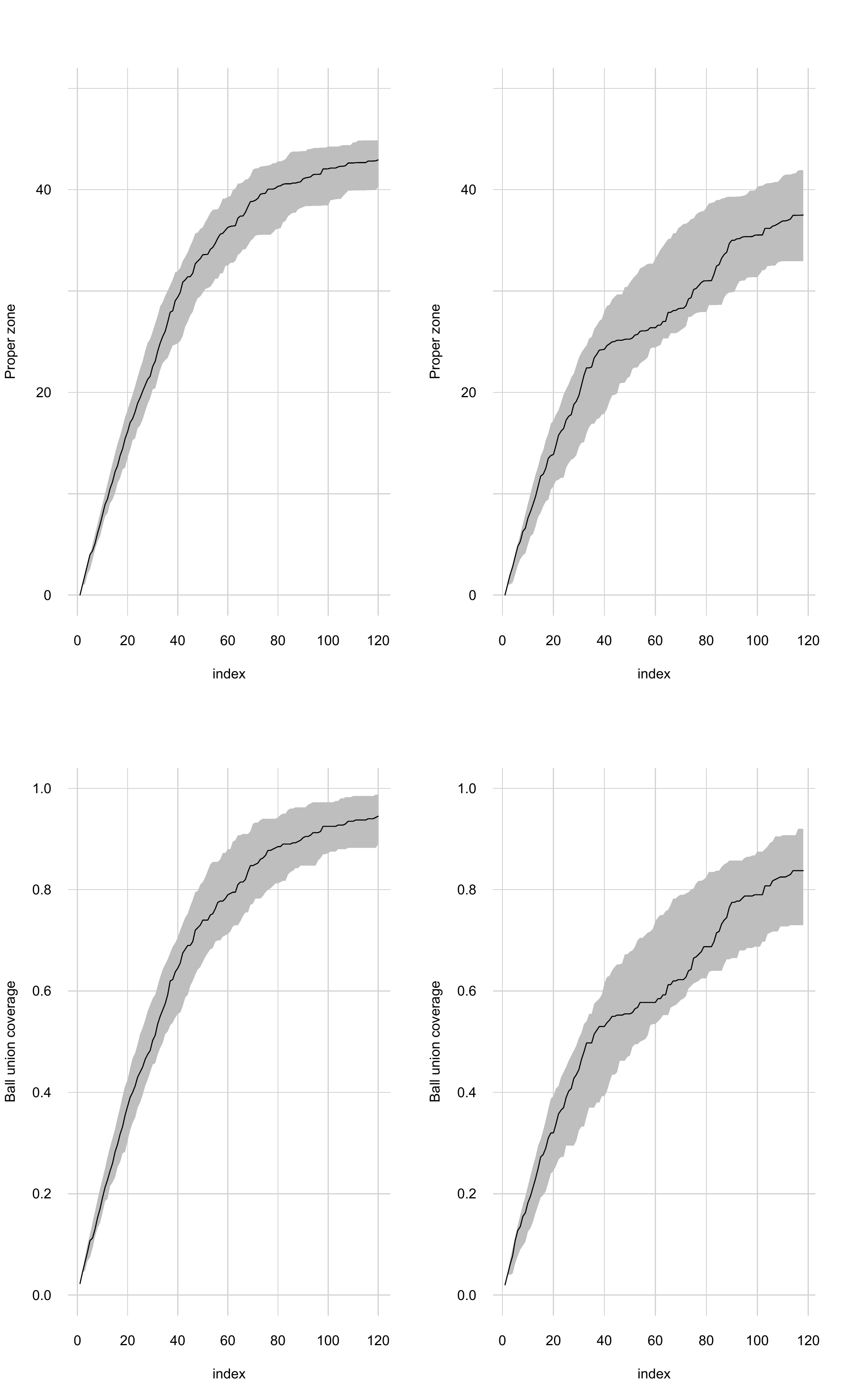}}
	\end{center}
	\caption{\small Summary statistics for Plot I (left column) and Plot II (right column).  Cumulative lagged clustering (first row), cumulative first contact distance (second row),
		cumulative proper zone statistic (third row) and the cumulative ball union coverage (last row). The black solid line stands for the data, and the grey area  represents pointwise envelopes estimated from 999 simulations of fitted model with parameters $(\hat{\theta}, \hat{r})=(0.17, 2.18)$ and  $(\hat{\theta}, \hat{r})=(0.65, 2.60)$ for Plot I and Plot II respectively.}
	\label{envplot}
\end{figure}

As can be seen in Figure \ref{envplot}, the four summary statistics estimated from the data stay within the simulated pointwise envelopes,  which indicates that the developed SSPP model fits well showing an agreement between the data and the fitted model. Also, we were able to catch the main spatial features of interest in  the forest dataset through the proposed summary statistics. On the other hand, the parameter estimates $\hat{\theta}$ and $\hat{r}$ are consistent with the exploratory study conducted by the $\LL$-function. For the Plot I, the obtained value of $\hat{\theta}$ is  small while it is large for Plot II. This indicates that the SSPP model reads the Plot I as a pattern where the new trees are  choosing non-occupied locations by the previous ones in a distance greater than 2.18, which   exhibits a spatial inhibition character of the pattern.  For Plot II, the new points favor the locations nearby the past locations occupied by the process with a high probability value of 0.65 and within a distance of 2.60, which is explained as a cause of  spatial clustering. 

By writing the self-interaction term $ \pp( y, \Sr(\XX_k, y))=\pp(r)$ as a function of the radius $r$ for both of the  plots, we get the piecewise constant function for Plot I as: 
\begin{linenomath} 
	\begin{equation*}
	\pp( r)=\begin{cases}
	0 & \text{if $ r < 0.2315$,} \\
	0.17 & \text{if $0.2315 \leq r < 2.18$,}\\
	0.83 & \text{if $ 2.18 \leq r, $}
	\end{cases}
	\end{equation*}
\end{linenomath}
and for Plot II as
\begin{linenomath} 
	\begin{equation*}
	\pp( r)=\begin{cases}
	0 & \text{if $ r < 0.4195$,} \\
	0.65 & \text{if $0.4195 \leq r < 2.60$,}\\
	0.35 & \text{if $ 2.60 \leq r. $}
	\end{cases}
	\end{equation*}
\end{linenomath}
The self-interaction behavior in the SSPP model can be extracted from the monotonicity of the cumulative lagged clustering function for both plots  shown by the dark solid line (Figure \ref{envplot}). In Plot II, the cumulative lagged clustering function is growing  faster to higher values than in Plot I, especially for the middle-sized trees (indexed from 40 to 80) where the cumulative ball union coverage area function also indicates a slowing behavior  meaning that there is no extension of the forest area inside $W$,  so these trees  choose locations in the neighborhood of the bigger trees. This was different for the small-sized trees that actually look for non-occupied zones inside the  window of observation, as shown by the cumulative ball union coverage area function. Also, the total coverage explains  the regular and the clustered nature of Plot I and Plot II, respectively, where it approaches the total area by trees in Plot I where smaller trees choose locations nearby older ones, while trees in Plot II cover only 80\% of the this total area.

On the other hand, the first contact distance function for Plot I shows a steady variation among all trees except for the first 20 trees where they avoid each others. In Plot II this statistic points out the behavior of the middle-sized trees where their first contact distances are shorter compared to other trees.    The cumulative proper zone statistic did not show much variability in terms of simulations, however, trees in Plot I are sharing resources at regular pace,  even when the coverage area is not extended that much anymore after the  time (order) 100. This was different in Plot II where the statistic shows a slight increase of the competition by middle-sized trees.

%    , while for the older treesthe curve tells that the trees are avoiding each other at a regular pace, even when the coverage area is not extended that much anymore after the  time (order) 43 where the trees covers almost 86\% of the area of the window, as the convex hull coverage curve shows.  The summary statistics related to the path length for both plots indicates a linear and stable behavior along the sequence with no specific large jumps between the locations of the trees.  

Comparing the model interpretation with the observed data in Figure \ref{plot}, we conclude that the developed SSPP model performs well by reading the self-interaction between the trees in the future and the past in both the clustered and the regular pattern,  and this self-interaction property,  that contains the memory and the learning through history, is well preserved under the SSPP model without loss of information.

\section{Conclusion}
\label{s:discuss}

%{\large{\texttt{discussion will be added after authors communications. Some general points can be:}}}

The sequential spatial point process introduced in this paper is a special case of the class of  spatio-temporal point processes where we work with ordered spatial points. In particular, this marginal process represents a bridge between the spatio-temporal point processes and the purely spatial point processes when a given order is adopted.

Our goal was to show that the mechanism of the sequential model represents a powerful modeling tool since it offers a causal description through successive conditioning, and revealing the information contained in the spatial memory. Moreover, self-interaction structure of the model enables interpreting the model parameters in terms of future-past dependence in a dynamic form. An additional advantage of the  model is that it is likelihood-based allowing to use well-defined statistical tools in analyzing data. However, evaluating the model requires constructing summary statistics that better describe the features of the phenomena in terms of order, which are seldom considered in the literature. For this, we used real data and we utilized  four summary statistics to evaluate the potential variations of data   assisted by Monte Carlo simulation. Since the functions are in terms of time/order, they allow us to  catch the variation of the feature in question. In our application, we were able to distinguish between  the variation exhibited by each class of tree sizes, large, medium-sized, and  small.    

Although the SSPP model was able to read patterns with different aggregations, the concept of the sequential model presented in this paper differs from other spatial point processes in the sense that it is  interpreting the structure of observed patterns beyond the general regular-clustered definition. It is reading their aggregation at a certain specific range given by the parameter $r$. In contrast one could fix this parameter in an approach similar to \citet{penttinen2016}, or more generally, propose a collection of suitable $r$ values representing the range of competition strength. This has an advantage in laser scanning application where the fixed $r$ could represent the crown radius of the tree, therefore the only parameter to be estimated would be $\theta$.

The order of the spatial dimension given in the paper emphasizes the dynamic evolution of the forest stand, as we believe that the trees with larger  sizes appear earlier than the small ones. Hence, our objective was to switch from the static point patterns to a dynamic one where the information that lies within the elevated spatial history is of great importance. However, other orders can be used when applying the sequential  spatial point processes which might depend on the context and on the observed phenomena. 

It is clear that our  model depends on the order used for the spatial points, i.e., when choosing different order, the interactions between a point and its past will definitely influence the density and the probabilistic properties of the process, except for the case of $\theta =0.5$ where the density is invariant under the choice of the order. This immediately reflects the non-exchangeability property of the model which usually does not exist in most of the structures of spatial point processes. The analysis introduced in this paper can be extended to compare different types of ordering  for static point patterns, which is a future task.

\section*{Acknowledgments}
%
%The authors thank Professor A. Sen for some helpful suggestions,
%Dr C. R. Rangarajan for a critical reading of the original version of the
%paper, and an anonymous referee for very useful comments that improved
%the presentation of the paper.\vspace*{-8pt}

Authors  would like to thank Juha Heikkinen, Mari Myllym\"{a}ki  (Natural Resources Institute Finland, Helsinki), and Kasper Kansanen (University of Eastern Finland) for useful remarks and discussions.  This work was funded by the Academy of Finland (Nb. 310073) 

%  If your paper refers to supplementary web material, then you MUST
%  include this section!!  See Instructions for Authors at the journal
%  website http://www.biometrics.tibs.org
%%%%%%%%%%%%%%%%%%%%%%%%%%%%%%%%%%%%%%%%
%%% Bibliography %%%%%%%%%%%%%%%%%%%%%%%
%%%%%%%%%%%%%%%%%%%%%%%%%%%%%%%%%%%%%%%%

%\section*{References}

\nocite{*}
\bibliographystyle{apalike}
\bibliography{manu}

\begin{thebibliography}{}

\bibitem[Baddeley et~al., 2015]{baddeley2015}
Baddeley, A., Rubak, E., and Turner, R. (2015).
\newblock {\em Spatial Point Patterns: Methodology and Applications with R}.
\newblock Chapman \& Hall/CRC.

\bibitem[Cressie, 1993]{cressie1993}
Cressie, N. A.~C. (1993).
\newblock {\em Statistics for Spatial Data}.
\newblock Wiley.

\bibitem[Daley and Vere-Jones, 2003]{daleyverejones2003}
Daley, D.~J. and Vere-Jones, D. (2003).
\newblock {\em An Introduction to the Theory of Point Processes. Volume I:
  Elementary Theory and Methods}.
\newblock New York: Springer-Verlag.

\bibitem[Daley and Vere-Jones, 2008]{daleyverejones2008}
Daley, D.~J. and Vere-Jones, D. (2008).
\newblock {\em An Introduction to the Theory of Point Processes. Volume II:
  General Theory and Structure}.
\newblock New York: Springer-Verlag.

\bibitem[Davison, 2008]{davison2008}
Davison, A.~C. (2008).
\newblock {\em Statistical Models}.
\newblock Cambridge University Press.

\bibitem[Diggle, 2013]{diggle2013}
Diggle, P.~J. (2013).
\newblock {\em Statistical Analysis of Spatial and Spatio-Temporal Point
  Patterns}.
\newblock Chapman \& Hall/CRC.

\bibitem[Diggle et~al., 1976]{diggle1976}
Diggle, P.~J., Besag, J., and Gleaves, J.~T. (1976).
\newblock Statistical analysis of spatial point patterns by means of distance
  methods.
\newblock {\em Biometrics}, pages 65--667.

\bibitem[Efron and Tibshirani, 1994]{efron1994}
Efron, B. and Tibshirani, R.~J. (1994).
\newblock {\em An Introduction to the Bootstrap}.
\newblock Chapman and Hall/CRC Press.

\bibitem[Evans, 1993]{evans1993}
Evans, J.~V. (1993).
\newblock Random and cooperative sequential adsorption.
\newblock {\em Reviews of Modern Physics}, 65:1281--1330.

\bibitem[Gablonsky and Kelley, 2001]{gablonsky2001}
Gablonsky, J. and Kelley, C. (2001).
\newblock A locally-biased form of the direct algorithm.
\newblock {\em Journal of Global Optimization}, 21:27--37.

\bibitem[Gonz\'{a}lez et~al., 2016]{gonzalez2016}
Gonz\'{a}lez, J.~A., Rodr\'{\i}guez-Cort\'{e}s, F.~J., Cronie, O., and Mateu,
  J. (2016).
\newblock Spatio-temporal point process statistics: A review.
\newblock {\em Spatial Statistics}, 18:505--544.

\bibitem[Illian et~al., 2008]{illian2008}
Illian, J., Penttinen, A., Stoyan, H., and Stoyan, D. (2008).
\newblock {\em Statistical Analysis and Modeling of Spatial Point Patterns}.
\newblock Wiley, Chichester, UK.

\bibitem[Jensen et~al., 2007]{jensen2007}
Jensen, E. B.~V., J\'{o}nsd\'{o}ttir, K.~Y., Schmiegel, J., and
  Barndorff-Nielsen, O.~E. (2007).
\newblock Spatio-temporal modelling with a view to biological growth.
\newblock {\em Statistical methods for spatio-temporal systems, Monographs on
  statistics and applied probability}, pages 47--75.

\bibitem[Johnson, 2010]{johnson2010}
Johnson, S.~G. (2010).
\newblock The nlopt nonlinear-optimization package.
\newblock {\em http://ab-initio.mit.edu/nlopt}.

\bibitem[Kansanen et~al., 2016]{kasper2016}
Kansanen, K., Vauhkonen, J., L\"{a}hivaara, T., and Meht\"{a}talo, L. (2016).
\newblock Stand density estimators based on individual tree detection and
  stochastic geometry.
\newblock {\em Canadian Journal of Forest Research}, 46:1359--1366.

\bibitem[Lieshout, 2006a]{lieshout2006a}
Lieshout, M. (2006a).
\newblock Campbell and moment measures for finite sequential spatial processes.
\newblock {\em Proceedings Prague Stochastics}, 48:215--224.

\bibitem[Lieshout, 2006b]{lieshout2006b}
Lieshout, M. (2006b).
\newblock Markovianity in space and time.
\newblock {\em IMS Lecture Notes--Monograph Series}, 48:154--168.

\bibitem[Lieshout, 2006c]{lieshout2006c}
Lieshout, M. (2006c).
\newblock Maximum likelihood estimation for random sequential adsorption.
\newblock {\em Advances in Applied Probability (SGSA)}, 38:889--898.

\bibitem[Lieshout and Capasso, 2009]{lieshout2009}
Lieshout, M. and Capasso, V. (2009).
\newblock Sequential spatial processes for image analysis.
\newblock {\em Electronic Transactions on Numerical Analysis}.

\bibitem[Lotwick and Silverman, 1982]{lotwick1982}
Lotwick, H.~W. and Silverman, B.~W. (1982).
\newblock Methods for analyzing spatial processes of several types of points.
\newblock {\em J. Roy. Stat. Soc. Series B}, 44:406--413.

\bibitem[Maltamo et~al., 2014]{maltamo2014}
Maltamo, M., Naesset, E., and Vauhkonen, J. (2014).
\newblock {\em Forestry Applications of Airborne Laser Scanning Concepts and
  Case Studies}.
\newblock Springer.

\bibitem[Meht\"{a}talo, 2006]{lauri2006}
Meht\"{a}talo, L. (2006).
\newblock Eliminating the effect of overlapping crowns from aerial inventory
  estimates.
\newblock {\em Canadian Journal of Forest Research}, 36:1649--1660.

\bibitem[M\o{}ller et~al., 2016]{moller2016}
M\o{}ller, J., Ghorbani, M., and Rubak, E. (2016).
\newblock Mechanistic spatio-temporal point process models for marked point
  processes, with a view to forest stand data.
\newblock {\em Biometrics}, 72:687--696.

\bibitem[M\o{}ller and Waagepetersen, 2004]{moller2004}
M\o{}ller, J. and Waagepetersen, R.~P. (2004).
\newblock {\em Statistical Inference and Simulation for Spatial Point
  Processes}.
\newblock Chapman \& Hall/CRC.

\bibitem[Mrkvi\u{c}ka et~al., 2018]{mrv2018}
Mrkvi\u{c}ka, T., Myllym\"{a}ki, M., Jilek, M., and Hahn, U. (2018).
\newblock A one-way anova test for functional data with graphical
  interpretation.
\newblock {\em ArXiv:1612.03608}.

\bibitem[Myllym\"{a}ki et~al., 2017]{mylly2017}
Myllym\"{a}ki, M., Mrkvi\u{c}ka, T., Grabarnik, P., Seijo, H., and Hahn, U.
  (2017).
\newblock Global envelope tests for spatial processes.
\newblock {\em Journal of the Royal Statistical Society Series B (Statistical
  Methodology)}, 79:381--404.

\bibitem[Packalen et~al., 2013]{packalen2013}
Packalen, P., Vauhkonen, J., Kallio, E., Peuhkurinen, J., Pitk\"{a}nen, J.,
  Pippuri, I., Strunk, J., and Maltamo, M. (2013).
\newblock Predicting the spatial pattern of trees by airborne laser scanning.
\newblock {\em International Journal of Remote Sensing}, 34(14):5154--5165.

\bibitem[Penttinen and Ylitalo, 2016]{penttinen2016}
Penttinen, A. and Ylitalo, A.~K. (2016).
\newblock Deducing self-interaction in eye movement data using sequential
  spatial point processes.
\newblock {\em Spatial Statistics}, 17:1--21.

\bibitem[Richardson and Kuester, 1973]{richardson1973}
Richardson, J.~A. and Kuester, J.~L. (1973).
\newblock The complex method for constrained optimization.
\newblock {\em Communications of the ACM}, 16:487--489.

\bibitem[Ripley, 1977]{ripley1977}
Ripley, B.~D. (1977).
\newblock Modeling spatial patterns.
\newblock {\em J. Roy. Stat. Soc. Series B}, 39:172--212.

\bibitem[Ripley, 1987]{ripley1987}
Ripley, B.~D. (1987).
\newblock {\em Stochastic simulation}.
\newblock Wiley.

\bibitem[Snyder and Miller, 1991]{synder}
Snyder, D.~L. and Miller, M.~I. (1991).
\newblock {\em Random Point Processes in Time and Space}.
\newblock Springer.

\bibitem[Stoyan et~al., 2017]{stoyan2017}
Stoyan, D., Rodr\'{\i}guez‐Cort\'{e}s, F.~J., Mateu, J., and Gille, W.
  (2017).
\newblock Mark variograms for spatio‐temporal point processes.
\newblock {\em Spatial Statistics}, 20:125--147.

\bibitem[Talbot et~al., 2000]{tablot}
Talbot, J., Tarjus, G., Van~Tassel, P.~R., and Viot, P. (2000).
\newblock From car parking to protein adsorption: An overview of sequential
  adsorption processes.
\newblock {\em Colloids and Surfaces A: Physicochemical and Engineering
  Aspects}, 165:28--324.

\bibitem[Vere-Jones, 2009]{verejones2009}
Vere-Jones, D. (2009).
\newblock Some models and procedures for space-time point processes.
\newblock {\em Environmental and Ecological Statistics}, 16:173--195.

\bibitem[Ylitalo, 2017]{ylitalo2017}
Ylitalo, A.~K. (2017).
\newblock {\em Statistical inference for eye movement sequences using spatial
  and spatio-temporal point processes}.
\newblock PhD thesis, University of Jyväskylä, Department of Mathematics and
  Statistics.
\newblock Report 160.

\end{thebibliography}

 \section*{Appendix}
The following tables are referenced in Section~\ref{app}. Tables contain summary of the estimated parameters of the 79 plots, associated with plots metrics. \vspace*{-8pt}

\begin{sidewaystable}
	\caption{\small First column indicates the plot identification number from 79 forest plots in Kiihtelysvaara. Columns 2-5 calculate the estimated parameters $\hat{\theta}$ and $\hat{r}$ and their associated bootstrap 95\% confidence intervals. Columns 6-10 contain characteristics of the plot as follows: The number of trees in the plot, the minimum, the maximum and the mean values of the diameters measured at breast height (DBH) of trees in the plot (cm). Last column shows  the size of the plot (m). Plot I and Plot II considered in the study are the plots  with identification number 5 and 74 respectively  }
	\label{t:two}
	\begin{center}\setlength{\tabcolsep}{6pt} 
		\def\~{\hphantom{0}}
		\scalebox{0.9}{
			\begin{tabular}{cccccccccc}
			%	\Hline
				{\small	Plot id.}	&
				\multicolumn{1}{c}{\small $\hat{\theta}$} & 
				\multicolumn{1}{c}{\small $95$\% CI for $\hat{\theta}$ } &
				\multicolumn{1}{c}{\small $\hat{r}$} & 
				\multicolumn{1}{c}{\small $95$\% CI for $\hat{r}$ } &
				\multicolumn{1}{c}{\small Nb. of indiv.} &
				\multicolumn{1}{c}{\small Min. DBH (cm)} &
				\multicolumn{1}{c}{\small Max. DBH (cm)}&
				\multicolumn{1}{c}{\small Mean DBH (cm)} &
				\multicolumn{1}{c}{\small Plot size (m $\times$ m)} 
			
				\\ 
				\hline
				
				$1$ & $0.20$ &  (0.11, 0.28)      & $3.80$ &  (3.69, 3.90)    & $114$ & $3.00$ & $44.50$ & $15.31$ & 30 x 30 \\ 
				$2$ & $0.30$ &  (0.25, 0.35)      & $3.25$ &  (3.07, 3.43)   & $73$ & $3.05$ & $34.00$ & $14.28$ & 25 x 25 \\ 
				$3$ & $0.23$ &  (0.17, 0.28)      & $2.42$ &  (2.23, 2.60)   & $54$ & $4.25$ & $29.00$ & $17.48$ & 25 x 25 \\ 
				$4$ & $0.39$ &  (0.27, 0.50)      & $3.13$ &  (2.11, 4.14)   & $100$ & $3.90$ & $26.30$ & $12.31$ & 25 x 25 \\ 
				$5$ & $0.17$ &  (0.13, 0.19)      & $2.18$ &  (2.12, 2.25)   & $120$ & $4.10$ & $23.15$ & $11.39$ & 25 x 25 \\ 
				$6$ & $0.34$ &  (0.31, 0.37)      & $3.15$ &  (2.57, 3.74)   & $104$ & $4.20$ & $50.10$ & $17.64$ & 30 x 30 \\ 
				$7$ & $0.75$ &  (0.69, 0.81)      & $0.50$ &  (0.35, 0.65)                & $104$ & $3.45$ & $36.75$ & $13.53$ & 30 x 30 \\ 
				$8$ & $0.11$ &  (0.06, 0.16)      & $1.20$ &  (1.13, 1.27)                 & $59$ & $2.50$ & $32.20$ & $17.62$ & 24 x 25 \\ 
				$9$ & $0.21$ &  (0.17, 0.25)      & $1.39$ &  (0.44, 2.32)     & $133$ & $3.85$ & $27.40$ & $13.91$ & 25 x 25 \\ 
				$10$ & $0.14$ &  (0.13, 0.15)      & $1.49$ &  (1.24, 1.74)       & $125$ & $3.95$ & $28.10$ & $13.78$ & 25 x 25 \\ 
				$11$ & $0.83$ &  (0.75, 0.91)      & $7.96$ &  (7.28, 8.64)              & $48$ & $3.30$ & $36.80$ & $17.20$ & 25 x 25 \\ 
				$12$ & $0.37$ &  (0.30, 0.44)      & $1.31$ &  (0.95, 1.68)                & $160$ & $2.30$ & $21.60$ & $9.25$ & 24 x 25 \\ 
				$13$ & $0.30$ &  (0.17, 0.43)      & $3.45$ &  (2.55, 4.35)  & $48$ & $4.95$ & $37.35$ & $17.38$ & 25 x 25 \\ 
				$14$ & $0.16$ &  (0.13, 0.19)      & $4.25$ &  (4.15, 4.35) & $34$ & $4.05$ & $39.60$ & $21.06$ & 25 x 25 \\ 
				$15$ & $0.14$ &  (0.10, 0.17)      & $1.19$ &  (1.15, 1.22) & $89$ & $3.05$ & $18.25$ & $11.50$ & 20 x 20 \\ 
				$16$ & $0.18$ &  (0.14, 0.22)      & $1.42$ &  (1.28, 1.55)  & $89$ & $3.45$ & $27.85$ & $13.49$ & 25 x 25 \\ 
				$17$ & $0.23$ &  (0.18, 0.28)      & $1.79$ &  (1.61, 1.97) & $61$ & $3.20$ & $22.20$ & $10.53$ & 20 x 20 \\ 
				$18$ & $0.09$ &  (0.07, 0.11)      & $1.49$ &  (1.45, 1.53)  & $77$ & $2.30$ & $21.95$ & $13.55$ & 20 x 20 \\ 
				$19$ & $0.88$ &  (0.84, 0.93)      & $6.78$ &  (6.19, 7,38) & $68$ & $5.15$ & $36.55$ & $16.86$ & 25 x 25 \\ 
				$20$ & $0.11$ &  (0.07, 0.15)      & $0.92$ &  (0.85, 0.98)  & $81$ & $2.50$ & $23.40$ & $13.63$ & 20 x 20 \\ 
				$21$ & $0.80$ &  (0.75, 0.85)      & $0.40$ &  (0.24, 0.55)  & $101$ & $2.85$ & $25.70$ & $10.83$ & 25 x 25 \\ 
				$22$ & $0.13$ &  (0.10, 0.16)      & $1.32$ &  (1.19, 1.45)  & $82$ & $2.95$ & $24.50$ & $14.49$ & 25 x 25 \\ 
				$23$ & $0.20$ &  (0.13, 0.27)      & $3.50$ &  (2.76, 4.23)  & $65$ & $2.10$ & $28.15$ & $13.49$ & 20 x 20 \\ 
				$24$ & $0.35$ &  (0.22, 0.48)      & $1.19$ &  (0.48, 1.90)  & $65$ & $4.25$ & $25.05$ & $12.72$ & 25 x 25 \\ 
				$25$ & $0.61$ &  (0.56, 0.66)      & $0.86$ &  (0.63, 1.09)  & $82$ & $2.65$ & $28.20$ & $10.61$ & 20 x 20 \\ 
				$26$ & $0.95$ &  (0.83, 1.06)      & $0.15$ &  (0.03, 0.27) & $72$ & $3.35$ & $24.20$ & $12.84$ & 25 x 25 \\ 
				$27$ & $0.23$ &  (0.19, 0.26)      & $2.61$ &  (2.51, 2.70) & $45$ & $3.00$ & $21.90$ & $15.03$ & 25 x 25 \\ 
				$28$ & $0.39$ &  (0.36, 0.42)      & $1.91$ &  (1.58, 2.24) & $58$ & $2.45$ & $22.65$ & $9.75$ & 20 x 20 \\ 
				$29$ & $0.80$ &  (0.75, 0.85)      & $0.55$ &  (0.26, 0.84) & $58$ & $3.00$ & $48.75$ & $12.40$ & 20 x 20 \\ 
				$30$ & $0.20$ &  (0.14, 0.25)      & $1.75$ &  (1.63, 1.87) & $47$ & $4.15$ & $29.00$ & $17.52$ & 25 x 25 \\ 
				
				\hline
		\end{tabular}}
	\end{center}
	%\end{table}
\end{sidewaystable}

\begin{sidewaystable}
	\caption{\small Contined. }
	\label{t:two}
	\begin{center}
		\scalebox{0.9}{
			\begin{tabular}{cccccccccc}
				%\Hline
				{\small	Plot id.}	&
				\multicolumn{1}{c}{\small $\hat{\theta}$} & 
				\multicolumn{1}{c}{\small $95$\% CI for $\hat{\theta}$ } &
				\multicolumn{1}{c}{\small $\hat{r}$} & 
				\multicolumn{1}{c}{\small $95$\% CI for $\hat{r}$ } &
				\multicolumn{1}{c}{\small Nb. of indiv.} &
				\multicolumn{1}{c}{\small Min. DBH (cm)} &
				\multicolumn{1}{c}{\small Max. DBH (cm)}&
				\multicolumn{1}{c}{\small Mean DBH (cm)} &
				\multicolumn{1}{c}{\small Plot size (m $\times$ m)} 
				\\ 
				\hline
				
				$31$ & $0.12$ &  (0.06, 0.17)      & $1.77$ &  (1.74, 1.80) & $51$ & $2.20$ & $25.55$ & $17.40$ & 25 x 25 \\ 
				$32$ & $0.25$ &  (0.15, 0.34)      & $3.90$ &  (3.08, 4.72) & $64$ & $3.10$ & $24.95$ & $13.97$ & 25 x 25 \\ 
				$33$ & $0.95$ &  (0.91, 0.99)      & $6.55$ &  (5.90, 7.20) & $85$ & $2.20$ & $28.75$ & $10.39$ & 20 x 20 \\ 
				$34$ & $0.33$ &  (0.31, 0.35)      & $2.32$ &  (1.92, 2.72) & $84$ & $1.95$ & $30.75$ & $13.48$ & 25 x 25 \\ 
				$35$ & $0.73$ &  (0.69, 0.76)      & $2.15$ &  (1.99, 2.31) & $71$ & $1.90$ & $23.80$ & $10.53$ & 20 x 20 \\ 
				$36$ & $0.24$ &  (0.12, 0.36)      & $4.90$ &  (3.46, 6.35) & $57$ & $2.30$ & $31.50$ & $14.26$ & 20 x 20 \\ 
				$37$ & $0.35$ &  (0.24, 0.46)      & $1.01$ &  (0.24, 1.78) & $50$ & $2.60$ & $33.25$ & $16.32$ & 25 x 25 \\ 
				$38$ & $0.25$ &  (0.18, 0.31)      & $2.89$ &  (2.51, 3.27) & $39$ & $2.65$ & $31.05$ & $17.49$ & 25 x 25 \\ 
				$39$ & $0.13$ &  (0.11, 0.15)      & $4.00$ &  (3.89, 4.12) & $75$ & $1.90$ & $37.70$ & $8.78$ & 19 x 20 \\ 
				$40$ & $0.13$ &  (0.09, 0.17)      & $3.23$ &  (3.16, 3.30) & $32$ & $13.10$ & $33.95$ & $21.84$ & 25 x 25 \\ 
				$41$ & $0.20$ &  (0.14, 0.26)      & $2.90$ &  (2.81, 2.99) & $38$ & $3.20$ & $25.55$ & $15.37$ & 25 x 25 \\ 
				$42$ & $0.39$ &  (0.29, 0.48)      & $0.87$ &  (0.19, 1.56) & $72$ & $5.15$ & $23.70$ & $14.25$ & 20 x 20 \\ 
				$43$ & $0.24$ &  (0.21, 0.27)      & $1.32$ &  (1.17, 1.48) & $90$ & $2.20$ & $21.50$ & $12.85$ & 20 x 20 \\ 
				$44$ & $0.17$ &  (0.13, 0.21)      & $0.99$ &  (0.87, 1.11) & $85$ & $2.50$ & $24.35$ & $11.47$ & 20 x 20 \\ 
				$45$ & $0.19$ &  (0.15, 0.23)      & $2.20$ &  (1.59, 2.82)  & $51$ & $3.95$ & $22.50$ & $15.85$ & 25 x 25 \\ 
				$46$ & $0.13$ &  (0.07, 0.20)      & $1.41$ &  (1.32, 1.50)  & $64$ & $3.00$ & $22.35$ & $13.73$ & 25 x 25 \\ 
				$47$ & $0.30$ &  (0.24, 0.36)      & $0.95$ &  (0.60, 1.29)  & $81$ & $1.90$ & $23.25$ & $10.06$ & 20 x 20 \\ 
				$48$ & $0.26$ &  (0.23, 0.30)      & $2.82$ &  (2.51, 3.13)  & $64$ & $2.05$ & $26.10$ & $15.43$ & 25 x 25 \\ 
				$49$ & $0.17$ &  (0.13, 0.20)      & $4.40$ &  (4.24, 4.55)  & $52$ & $2.45$ & $28.40$ & $12.40$ & 25 x 25 \\ 
				$50$ & $0.62$ &  (0.59, 0.66)      & $1.27$ &  (0.97, 1.58)  & $113$ & $2.95$ & $44.05$ & $15.66$ & 30 x 30 \\ 
				$51$ & $0.12$ &  (0.10, 0.15)      & $2.69$ &  (2.60, 2.77)  & $42$ & $5.90$ & $40.80$ & $28.40$ & 30 x 30 \\ 
				$52$ & $0.12$ &  (0.08, 0.16)      & $2.00$ &  (1.18, 2.17)  & $51$ & $5.05$ & $38.65$ & $25.78$ & 30 x 30 \\ 
				$53$ & $0.14$ &  (0.11, 0.16)      & $2.49$ &  (2.42, 2.56) & $49$ & $2.15$ & $40.75$ & $26.44$ & 30 x 30 \\ 
				$54$ & $0.09$ &  (0.06, 0.13)      & $2.23$ &  (2.16, 2.31) & $40$ & $4.30$ & $37.80$ & $18.50$ & 25 x 25 \\ 
				$55$ & $0.14$ &  (0.11, 0.16)      & $3.60$ &  (3.47, 3.72) & $46$ & $3.20$ & $34.80$ & $18.37$ & 25 x 25 \\ 
				$56$ & $0.21$ &  (0.15, 0.28)      & $1.47$ &  (1.31, 1.63) & $70$ & $2.40$ & $48.65$ & $17.36$ & 30 x 30 \\ 
				$57$ & $0.33$ &  (0.30, 0.36)      & $3.05$ &  (2.61, 3.50) & $58$ & $2.10$ & $32.80$ & $18.65$ & 30 x 30 \\ 
				$58$ & $0.06$ &  (0.03, 0.08)      & $1.87$ &  (1.82, 1.92) & $45$ & $2.35$ & $26.95$ & $17.82$ & 25 x 25 \\ 
				$59$ & $0.21$ &  (0.18, 0.24)      & $2.83$ &  (2.32, 3.33) & $54$ & $2.00$ & $35.10$ & $13.59$ & 25 x 25 \\ 
				$60$ & $0.21$ &  (0.19, 0.23)      & $2.21$ &  (1.81, 2.61) & $50$ & $4.40$ & $43.00$ & $17.60$ & 25 x 25 \\ 
				\hline
		\end{tabular}}
	\end{center}
	%\end{table}
\end{sidewaystable}

\begin{sidewaystable}
	\caption{\small Continued }
	\label{t:two}
	\begin{center}
		\scalebox{0.9}{
			\begin{tabular}{cccccccccc}
			%	\Hline
				{\small	Plot id.}	&
				\multicolumn{1}{c}{\small $\hat{\theta}$} & 
				\multicolumn{1}{c}{\small $95$\% CI for $\hat{\theta}$ } &
				\multicolumn{1}{c}{\small $\hat{r}$} & 
				\multicolumn{1}{c}{\small $95$\% CI for $\hat{r}$ } &
				\multicolumn{1}{c}{\small Nb. of indiv.} &
				\multicolumn{1}{c}{\small Min. DBH (cm)} &
				\multicolumn{1}{c}{\small Max. DBH (cm)}&
				\multicolumn{1}{c}{\small Mean DBH (cm)} &
				\multicolumn{1}{c}{\small Plot size (m $\times$ m)} 
				\\ 
				\hline

				$61$ & $0.33$ &  (0.30, 0.36)      & $3.15$ &  (2.54, 3.76) & $56$ & $3.60$ & $36.65$ & $17.29$ & 30 x 30 \\ 
				$62$ & $0.28$ &  (0.25, 0.32)      & $2.23$ &  (2.00, 2.46) & $79$ & $3.40$ & $31.75$ & $16.62$ & 30 x 30 \\ 
				$63$ & $0.21$ &  (0.18, 0.23)      & $3.00$ &  (2.86, 3.15) & $59$ & $4.35$ & $37.20$ & $19.72$ & 30 x 30 \\ 
				$64$ & $0.25$ &  (0.19, 0.31)      & $3.55$ &  (3.31, 3.78) & $81$ & $3.75$ & $40.00$ & $12.38$ & 25 x 25 \\ 
				$65$ & $0.22$ &  (0.20, 0.25)      & $1.80$ &  (1.70, 1.90) & $101$ & $3.35$ & $40.45$ & $14.67$ & 25 x 25 \\ 
				$66$ & $0.22$ &  (0.20.0.25)       & $2.11$ &  (1.75, 2.48) & $64$ & $3.35$ & $33.65$ & $18.41$ & 30 x 30 \\ 
				$67$ & $0.25$ &  (0.21, 0.29)      & $2.45$ &  (2.24, 2.66) & $57$ & $4.85$ & $34.10$ & $20.88$ & 30 x 30 \\ 
				$68$ & $0.17$ &  (0.14, 0.20)      & $2.00$ &  (1.59, 2.41) & $60$ & $2.30$ & $31.70$ & $18.36$ & 25 x 25 \\ 
				$69$ & $0.28$ &  (0.21, 0.34)      & $2.19$ &  (1.61, 2.77) & $49$ & $3.75$ & $36.25$ & $21.97$ & 30 x 30 \\ 
				$70$ & $0.35$ &  (0.30, 0.40)      & $1.08$ &  (0.69, 1.46) & $85$ & $4.05$ & $30.65$ & $12.08$ & 20 x 20 \\ 
				$71$ & $0.33$ &  (0.23, 0.43)      & $3.90$ &  (2.98, 4.82) & $115$ & $4.25$ & $26.70$ & $10.78$ & 20 x 20 \\ 
				$72$ & $0.11$ &  (0.08, 0.15)      & $2.05$ &  (1.96, 2.15) & $33$ & $4.90$ & $27.55$ & $19.05$ & 20 x 20 \\ 
				$73$ & $0.31$ &  (0.18, 0.43)      & $2.50$ &  (1.72, 3.27) & $55$ & $3.35$ & $27.10$ & $12.87$ & 20 x 20 \\ 
				$74$ & $0.65$ &  (0.58, 0.74)      & $2.60$ &  (1.06, 4.13) & $118$ & $2.30$ & $41.95$ & $12.11$ & 30 x 30 \\ 
				$75$ & $0.33$ &  (0.26, 0.41)      & $3.90$ &  (3.02, 4.78) & $113$ & $1.60$ & $39.40$ & $8.14$ & 25 x 25 \\ 
				$76$ & $0.65$ &  (0.60, 0.70)      & $1.20$ &  (0.87, 1.53) & $105$ & $2.65$ & $28.75$ & $14.13$ & 25 x 25 \\ 
				$77$ & $0.22$ &  (0.19, 0.26)      & $5.87$ &  (5.77, 5.96) & $84$ & $2.85$ & $49.10$ & $13.38$ & 25 x 25 \\ 
				$78$ & $0.80$ &  (0.77, 0.83)      & $0.50$ &  (0.44, 0.56)   & $100$ & $2.70$ & $39.15$ & $9.37$ & 25 x 25 \\ 
				$79$ & $0.30$ &  (0.26, 0.33)      & $1.00$ &  (0.82, 1.18) & $94$ & $3.20$ & $19.45$ & $10.95$ & 25 x 25 \\ 
				\hline
		\end{tabular}}
	\end{center}
	%\end{table}
\end{sidewaystable}

\end{document}